%% file: current_nat_sci_rep/main.tex
\documentclass[fleqn,10pt]{wlscirep}
\usepackage[utf8]{inputenc}
\usepackage[T1]{fontenc}
\usepackage{diagbox}  %
\usepackage{array}    %
\usepackage{lineno}
\title{High contrast holography through dual modulation}

\author[1, *]{Leyla Kabuli}
\author[1]{Oliver Cossairt}
\author[1]{Florian Schiffers}
\author[1]{Nathan Matsuda}
\author[1]{Grace Kuo}
\affil[1]{Reality Labs Research, Meta, USA}
\affil[*]{Corresponding author. Email: lakabuli@berkeley.edu}

\begin{abstract} %
Holographic displays are a promising technology for immersive visual experiences, and their potential for compact form factor makes them a strong candidate for head-mounted displays. However, at the short propagation distances needed for a compact, head-mounted architecture, image contrast is low when using a traditional phase-only spatial light modulator (SLM). Although a complex SLM could restore contrast, these modulators require bulky lenses to optically co-locate the amplitude and phase components, making them poorly suited for a compact head-mounted design. In this work, we introduce a novel architecture to improve contrast: by adding a low resolution amplitude SLM a short distance away from the phase modulator, we demonstrate peak signal-to-noise ratio improvement up to 31 dB in simulation and 6.5 dB experimentally compared to phase-only modulation, even when the amplitude modulator is 60$\times$ lower resolution than its phase counterpart. We analyze the relationship between diffraction angle and amplitude modulator pixel size, and validate the concept with a benchtop experimental prototype. By showing that low resolution modulation is sufficient to improve contrast, we open new design spaces for high-contrast holographic displays.
\end{abstract}
\begin{document}

\flushbottom
\maketitle
\thispagestyle{empty}

\section*{Introduction}

\input{current_nat_sci_rep/1_introduction}

\section*{Results}

\input{current_nat_sci_rep/2_results}

\section*{Discussion}
\input{current_nat_sci_rep/3_discussion}

\section*{Methods}

\input{current_nat_sci_rep/4_materials_and_methods}

\section*{Data availability}
Additional data supporting the results in this paper are available from the corresponding author upon reasonable request.

\bibliography{current_nat_sci_rep/main}

\section*{Acknowledgments}

Flower source image by Paul Longinidis (CC BY 2.0) is available at~\url{https://www.flickr.com/photos/128235612@N06/18033751041/}. Image was cropped but otherwise unmodified.
Houses source image by Madeleine Deaton (CC BY 2.0). This research was supported by Meta.

\section*{Author contributions}
L.K. and G.K. developed the concept. L.K. designed and built the experimental setup with input from G.K. and N.M.. L.K. captured experimental data and performed simulations with input from G.K., O.C., and N.M.. L.K., O.C., and F.S. implemented the system calibration with input from G.K. and N.M.. G.K., N.M., and O.C. supervised the work. All authors contributed to the writing and review of the manuscript and Supplementary Material.

\section*{Competing interests}
The authors declare no competing interests.

\newcommand{\beginsupplement}{%
        \setcounter{table}{0}
        \renewcommand{\thetable}{S\arabic{table}}%
        \setcounter{figure}{0}
        \renewcommand{\thefigure}{S\arabic{figure}}%
        \setcounter{section}{0}
        \renewcommand{\thesection}{S\arabic{section}}%
        \setcounter{equation}{0}
        \renewcommand{\theequation}{S\arabic{equation}}
     }

\newpage 
\beginsupplement

\setcounter{page}{14}

\section*{\fontsize{20}{25}\selectfont High contrast holography through dual modulation: \\
Supplementary material}

\vskip8pt

\section*{\fontsize{12}{16}\selectfont 
Leyla~Kabuli$^{1, \ast}$, 
Oliver~Cossairt$^1$,
Florian~Schiffers$^1$,
Nathan~Matsuda$^1$,
Grace~Kuo$^1$ }

\small$^1$ Reality Labs Research, Meta, USA \\
\small$^\ast$ Corresponding author. Email: lakabuli@berkeley.edu 

\vskip18pt

\noindent
This supplementary material includes additional implementation details, simulation analysis, examples, and experimental results.

\input{current_nat_sci_rep/6_supplement_text}

\end{document}

%% file: current_nat_sci_rep/1_introduction.tex
\noindent
Holographic displays can create imagery with accurate 3D focal cues, ocular parallax, and the ability to compensate for users' glasses prescriptions entirely in software.
These displays also have the potential to be very compact, making them particularly well-suited for head-mounted applications, such as virtual and augmented reality \cite{dpacmaimone17}.

To form an image, holographic displays use a spatial light modulator (SLM), which shapes the incident beam into an image through diffraction.
The most commonly used SLMs modulate only the phase of light, which is not directly visible to the human eye.
Instead, the SLM bends light locally, and after some propagation distance, light is concentrated (or spread out) enabling interference between different parts of the beam; the interference creates intensity differences that form the image.
However, as we'll discuss later, the finite pixel size on the SLM limits the maximum diffraction angle.
Therefore, if the propagation distance between the SLM and image plane is too short, light cannot be moved across the whole image plane and contrast suffers.

This is especially relevant for head-mounted displays (HMDs) where form factor is particularly important.
Consider the HMD architecture where the headset track length is minimized by placing the SLM directly against the eyepiece~\cite{bunnyears22}.
To produce an image at optical infinity, the propagation distance from the SLM to the image plane should be equal to the eyepiece focal length, typically 20 - 30~mm for a virtual reality headset.
However, as we'll show in this work, with commercially available SLMs, this propagation distance is too short to generate high-contrast images with phase-only modulation.

Complex modulation can overcome this issue since intensity is directly varied at the SLM plane without requiring any propagation.
However, complex SLMs with compact form factor are not currently available.
Although there are approaches to generate complex modulation from a single phase or amplitude SLM (e.g. double phase amplitude encoding~\cite{ogdpac78} and single-sideband filtering~\cite{singlesideband68}), these require bulky 4$f$ systems to filter in the Fourier plane, making them poorly suited for HMDs.
Other approaches have similar challenges: for example, using a 4$f$ relay system to optically co-locate separate phase and amplitude SLMs also sacrifices form factor~\cite{twoslmsphasefirst07}.
Finally, amplitude-only modulation cannot create high-quality holograms without a filter in the Fourier plane due to the twin-image problem~\cite{amptwinimage11}.

\begin{figure}[!t]
\includegraphics[width=\textwidth]{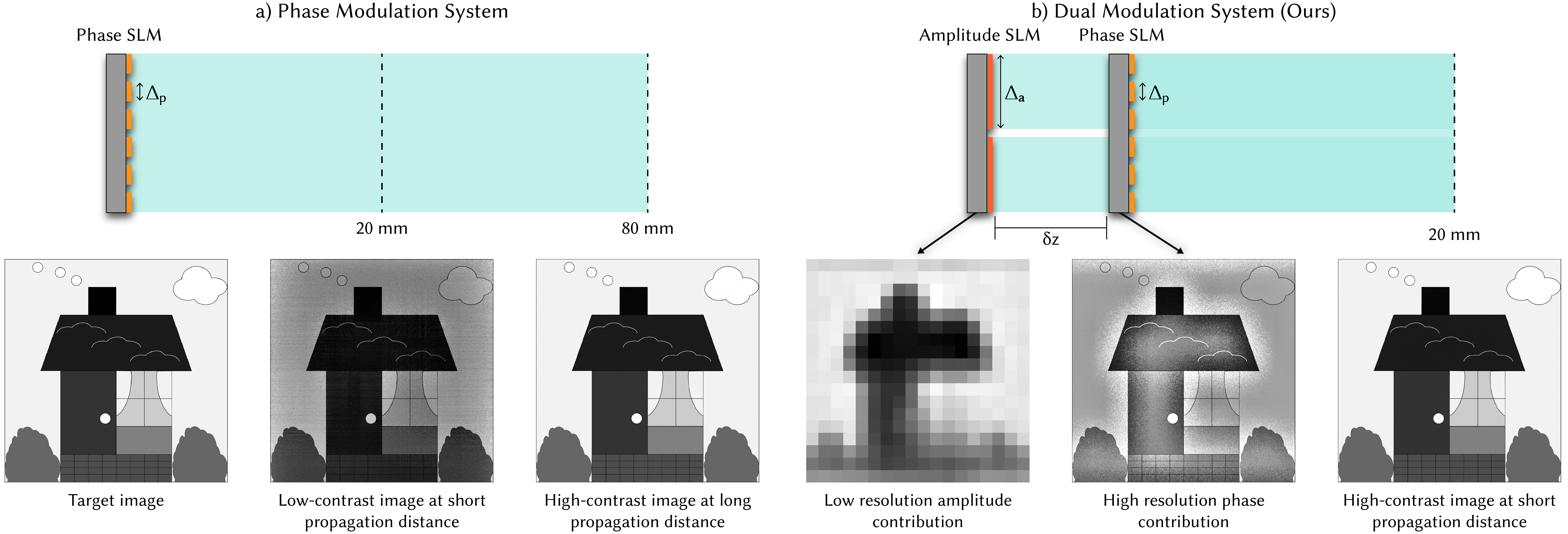}
  \caption{\textbf{Dual modulation system architecture.} Dual modulation holography combines a low resolution amplitude spatial light modulator (SLM) with a phase SLM to produce high-contrast images at short propagation distances. 
  (\textbf{a}) Traditional phase-only modulation produces images with low contrast at short propagation distances (20~mm). These images have intensity errors and variations across large uniform regions. Increasing the propagation distance (80~mm) can improve contrast.
  (\textbf{b}) Our dual modulation system places a low resolution amplitude SLM at a small distance ($\delta z$) in front of or behind a phase SLM in order to produce images with high contrast at short propagation distances (20~mm). The amplitude SLM contributes to low-frequency and uniform intensity regions, while the phase SLM prioritizes high-frequency details and transitions between light and dark regions.
  }
  \label{fig:overview_and_contrast}
\end{figure}

Our insight is that improved contrast does not require full complex modulation. 
An ideal complex modulator has co-located control of phase and amplitude at every pixel, but we can relax these constraints to be more practical if we primarily care about contrast.
We propose a new architecture, shown in Fig.~\ref{fig:overview_and_contrast}, which uses two modulators, phase and amplitude, but allows for a small gap between them.
This eliminates the need for an optical relay to co-locate the SLMs;
instead, the SLMs can be physically placed next to each other, paving the way to a compact device.

However, in this compact configuration at least one of the SLMs must be transmissive. Transmissive amplitude modulators are common in liquid crystal (LC) devices but tend to have larger pixel sizes than their reflective counterparts. This size difference is driven by fill factor: in a reflective device, electronics can be put behind the pixels, leading to fill factors of over 90\%, even for small pixels under 4~\textmu m~\cite{holoeyegaea}. However, with transmissive modulators, the electronics are always in the beam path, so larger pixel sizes are required to improve fill factor~\cite{curatu2009analysis}.

In this work, we demonstrate that contrast can be dramatically improved \textit{even when the amplitude modulator is much lower resolution than the phase modulator}. We use diffraction theory to recommend a pixel size for the amplitude modulator, and we show that despite the coarse pixel pitch, we can still create high resolution imagery and defocus cues. Since our approach uses two SLMs (low resolution amplitude and high resolution phase), we refer to this architecture as ``dual modulation.'' Dual modulation has most of the benefits of complex modulation but is significantly more flexible, enabling new designs for future display systems.

%% file: current_nat_sci_rep/2_results.tex
To form a hologram, an SLM modulates the electric field of an incident coherent illumination source, as shown in Fig.~\ref{fig:overview_and_contrast}a.
The resulting modulated electric field propagates in free space, and its intensity is captured by a camera sensor or human eye.
For an SLM with modulation pattern $\mathbf{m}(\vec{x})$ of 2D spatial coordinate $\vec{x}$, the image formation model is 
\begin{equation}
\begin{aligned}
    \mathbf{G}_z(\vec{x}) &= \mathcal{A}_z\bigl\{\mathbf{G}_0(\vec{x}) \odot \mathbf{m}(\vec{x})\bigr\}, \\
    \mathbf{I}_z(\vec{x}) &= |\mathbf{G}_z(\vec{x})|^2,
    \label{eq:phase_modulation_model}
\end{aligned}
\end{equation}
where $\mathbf{G}_z(\cdot)$ and $\mathbf{I}_z(\cdot)$ are the electric field and the intensity at propagation distance $z$ from the SLM respectively, and $\odot$ denotes pointwise multiplication.
$\mathbf{G}_0(\cdot)$ is the incident electric field on the SLM, typically a plane wave with uniform energy, $\mathbf{G}_0(\vec{x}) = 1$.
$\mathcal{A}_z$ is the angular spectrum method (ASM) propagation operator as detailed in Methods.

\subsection*{Contrast and light displacement}
\label{sec:angle-of-displacement}

\begin{figure}[t]
\includegraphics[width=\textwidth]{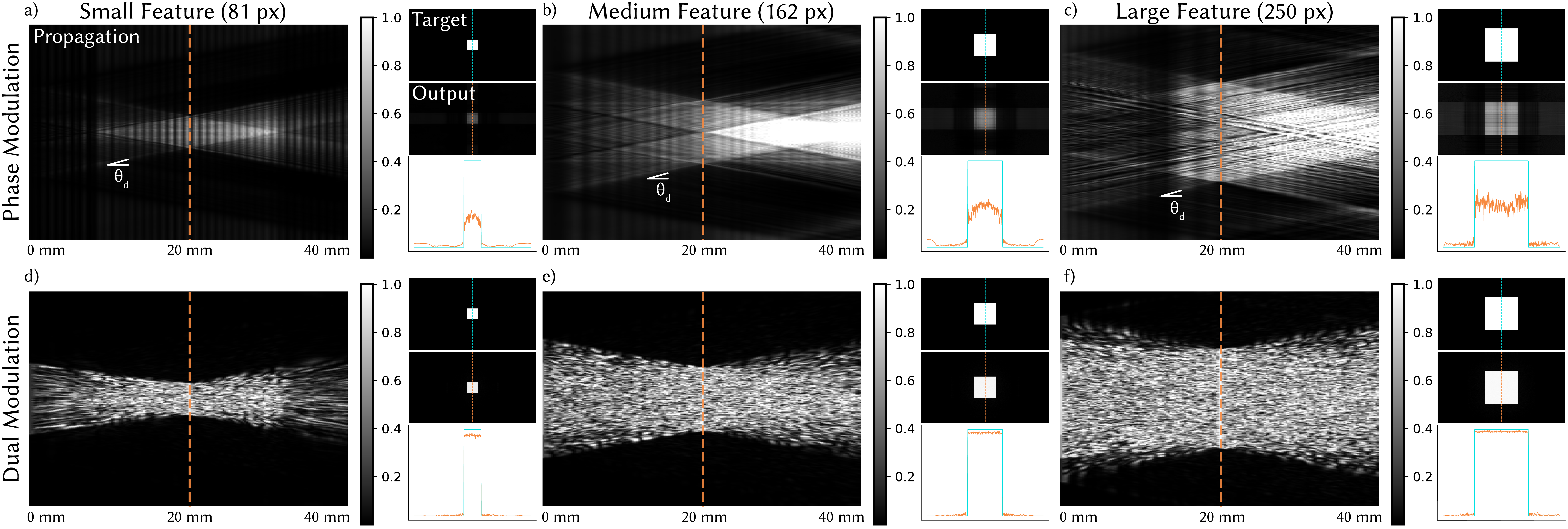}
  \caption{\textbf{Contrast in holographic displays.}
We simulate holograms where the target image is a white square of varying size. We visualize the intensity cross-section as a function of propagation distance, noting that light cannot be bent beyond $\theta_d$. With phase-only modulation (\textbf{a-c}), light at the edges of the SLM cannot reach the white square and therefore adds background intensity and reduces contrast. In addition, for the large square size (\textbf{c}), light cannot be moved from outside the square to the center, resulting in undesirable intensity oscillations. Finally, phase-only modulation creates an unnatural bright spot at longer propagation distances, which is not desirable if natural defocus cues are needed. In comparison, with our dual modulation approach (\textbf{d-f}), light that cannot be moved into the square is blocked at the amplitude SLM, which improves contrast and uniformity over all square sizes and creates natural defocus. %
}
  \label{fig:angle_limits}
\end{figure}

Ideally, the SLM would modulate both the phase and amplitude of the electric field, but complex SLMs are not available with compact form factors.
Instead, the most commonly-used SLMs modulate only phase, which locally bends light at the SLM plane.
Although this modulation is not immediately visible, after propagating to the image plane, light is concentrated in some areas and displaced in others to form bright and dark regions.

Intuitively, light only needs to be displaced by a small distance to create high-frequency details, but more displacement is needed to create larger, low-frequency features.
This trend mimics the relationship found in phase imaging using the Transport of Intensity Equation. In phase retrieval, short propagation distances best recover high-frequency details, whereas long propagation distances increase sensitivity to low-frequency features~\cite{lauratie14}.

When forming a hologram, to have full control over the image, ideally light could be moved from any position on the SLM to any position in the image plane, enabling interference between any two parts of the incident beam.
However, the SLM cannot move light arbitrarily far.
The maximum lateral displacement, $d$, depends on the maximum diffraction angle at the SLM, $\theta_d$, and the propagation distance, $z$, as follows:
\begin{equation}
\begin{aligned}
    d &= z \tan \theta_d, \\
    \theta_d &= \sin^{-1} \left(\frac{\lambda}{2 \Delta_p} \right).
\label{eq:maximum_phase_displacement}
\end{aligned}
\end{equation}
Here, the maximum diffraction angle is determined by the pixel pitch, $\Delta_p$, of the SLM, where smaller pitches enable more diffraction.

To make this more concrete, let's consider specific parameters compatible with a near-eye display architecture:
$\Delta_p = 8$~\textmu m, $z = 20$~mm,  $\lambda = 520$~nm.
In this scenario, the maximum displacement, $d$, is 650~\textmu m or 81.3 pixels.
Put in context, for an SLM with $1920 \times 1080$ pixels, the light can't be moved more than 5\% of the total horizontal field of view (FOV).
Although smaller pixel pitch or longer propagation distance could increase $d$, changing these parameters negatively affects FOV and form factor, respectively~\cite{bunnyears22}.

To demonstrate the relationship between $d$ and contrast, Fig.~\ref{fig:angle_limits} shows simulations of holograms where the target image is a white square of varying size. %
As expected, the light cannot propagate at an angle greater than $\theta_d$, so only light immediately surrounding the square can be shifted into the bright region.
The optimized SLM pattern leaves the rest of the FOV unmodulated, which adds background intensity to the black regions, reducing contrast.
Contrast is further reduced in the case where the white square is larger than $2d$ (Fig.~\ref{fig:angle_limits}c);
here, the SLM cannot shift light from outside the square to the center, so we see reduced brightness in the middle of the square and additional intensity oscillations as the optimizer attempts to make the square as uniform as possible.

This example highlights how phase-only modulation creates contrast by redirecting light from dark to bright regions, but when these regions are too far apart, full contrast cannot be achieved.
We note that approaches like temporal multiplexing can reduce speckle and improve image quality but cannot diffract light further than a single frame and therefore suffer the same contrast limitations.
Complex modulation can improve contrast by blocking light at the SLM, enabling dark regions even if they are not next to something bright.
However, complex modulation is not currently a practical option for HMDs since most configurations require a bulky 4$f$ system.
\textit{Our goal is to maintain the benefits of complex modulation but with a realistic path towards a compact system.}

\subsection*{Dual modulation with low resolution amplitude}
We introduce a dual modulation approach, shown in Fig.~\ref{fig:overview_and_contrast}b, where, like a true complex SLM, we propose combining phase and amplitude modulators for better light control.
However, to make the system more practical, we propose relaxing the requirements of a complex SLM in two ways:
(1) we allow for a gap between the phase and amplitude components, eliminating the need for a 4$f$ relay system between SLMs, and (2) we allow the resolution of the amplitude component to be significantly lower than the phase component.
Lower resolution is more compatible with transmissive SLM technology since larger pixels can have better fill factor.
In turn, transmissive SLMs provide more design flexibility when targeting a compact form factor. 
Even though this architecture has fewer degrees of freedom compared to a complex SLM, we'll show that it's sufficient to produce high contrast images.

We can model our dual modulation system as 
\begin{equation}
\begin{aligned}
    \mathbf{G}_z(\vec{x}) &= \mathcal{A}_z \Big\{ 
    \mathcal{A}_{\delta z} \big \{
    \mathbf{G}_0(\vec{x}) \odot \mathbf{m_a}(\vec{x}) \big \}
    \odot \mathbf{m_p}(\vec{x}) \Big \}, \\
    \mathbf{I}_z(\vec{x}) &= |\mathbf{G}_z(\vec{x})|^2,
\label{eq:dual_modulation_model}
\end{aligned}
\end{equation}
where $\mathbf{m_a}(\cdot)$ is the amplitude modulation,  $\mathbf{m_p}(\cdot)$ is the phase modulation, and $\delta z$ is the distance between the SLMs. This represents the scenario where the amplitude modulator is before the phase modulator; however, the SLM order can easily be inverted by swapping $\mathbf{m_a}$ and $\mathbf{m_p}$ in Eq. \eqref{eq:dual_modulation_model}.

\subsubsection*{Amplitude pixel size}
Next, we consider what amplitude pixel size is needed for high contrast.
Within the area subtended by a single amplitude pixel, image formation is essentially the same as with phase-only modulation, i.e. the total intensity can be controlled, but the image is formed by locally bending light.
Based on the analysis above, %
we know we have complete control of the image if light can be moved from any point on the SLM to interfere with light at any point in the image plane.
With the low resolution amplitude modulator, we've effectively split the FOV into small patches, so the necessary condition is
\begin{equation}
    \Delta_a \leq d,
\label{eq:maximum_pixel_size}
\end{equation}
where $d$ is defined in Eq.~\eqref{eq:maximum_phase_displacement}.

We note that further shrinking the amplitude pixels will continue to provide benefit by increasing the number of degrees of freedom in the system, but smaller pixels are not needed to enhance contrast. 
On the other hand, larger values of $\Delta_a$ that do not meet Eq.~\eqref{eq:maximum_pixel_size} can sometimes yield high-quality images, similar to how a phase-only modulator may be sufficient for favorable content.
However, with $\Delta_a$ above the bound in Eq.~\eqref{eq:maximum_pixel_size}, the SLM may not be able to shift light far enough for arbitrary images, which could result in artifacts at the boundaries between amplitude pixels.

\subsubsection*{Impact of dual modulation}
Figure~\ref{fig:angle_limits}d-f demonstrates the impact of our dual modulation scheme. Once again, the target image is a white square of varying size, and we simulate an amplitude SLM with $\Delta_a = 480$~\textmu m, which meets the condition in Eq.~\eqref{eq:maximum_pixel_size}. The amplitude modulator blocks light that is too far away from the white square ($>d$) to contribute, therefore eliminating background intensity and greatly improving contrast over all square sizes.

In addition, the intensity of the light as a function of $z$ is more natural with dual modulation. Assuming we want the image plane to resemble a traditional display with a diffuse surface, we expect equivalent defocus for $\pm z$. However, with phase-only modulation (Fig.~\ref{fig:angle_limits}a-c), bright parts of the image tend to grow in size when $z$ decreases and to shrink in size when $z$ increases, eventually creating an unnaturally bright spot at larger $z$. In comparison, holograms with dual modulation defocus uniformly as the propagation distance moves away from the image plane.

Finally, to highlight the benefit of dual modulation compared to true complex modulation, we consider specifications for a transmissive SLM. 
The inactive area of a transmissive SLM can be as low as a 2.8~\textmu m border~\cite{curatu2009analysis}.
For true complex modulation with an 8~\textmu m pixel pitch, this yields a fill factor of only 42\% which will substantially decrease diffraction efficiency.
However, with the larger 480~\textmu m pixels used in Fig.~\ref{fig:angle_limits} and the same border width, the fill factor is 98.8\%, higher than many reflective SLMs used regularly for holography~\cite{holoeyegaea}. We analyze the effect of fill factor on image quality in the Supplementary Material (Fig. S4) and show that the effects of the border are negligible with the proposed large pixel sizes.

\subsection*{Dual modulation simulations}

\begin{figure}
\includegraphics[width=\textwidth]{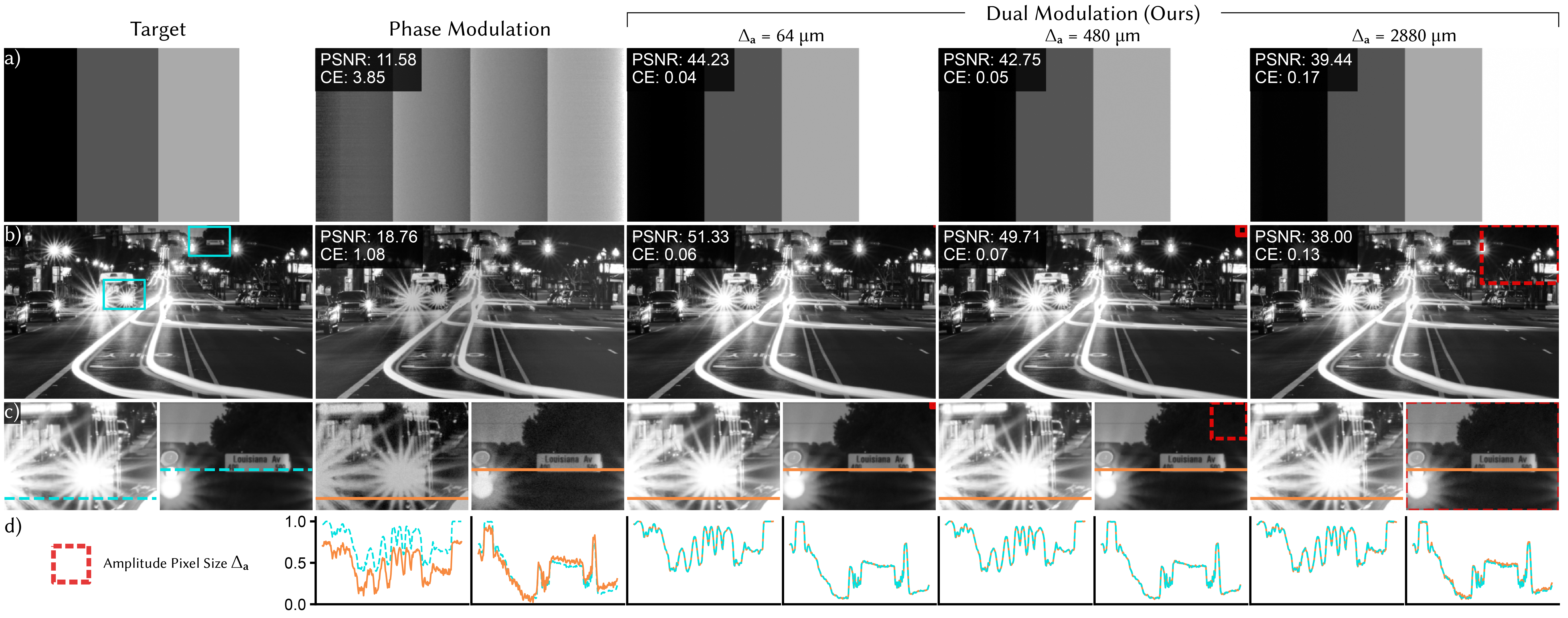}
  \caption{\textbf{Dual modulation (simulation).}
(\textbf{a}) Even in an ideal simulation, phase-only modulation has difficulty creating large uniform regions, like in this image of vertical bars. Our dual modulation approach can correctly create the target image even when the amplitude pixel size ($\Delta_a$) is large. (\textbf{b}) Dual modulation also improves contrast on a natural scene, although in this case there is some degradation in PSNR for the largest value of $\Delta_a$, which is outside the recommended bound in Eq.~\eqref{eq:maximum_pixel_size}. (\textbf{c}) Although the amplitude SLM is low resolution, fine details in the image are correctly created even when they are significantly smaller than $\Delta_a$, which is depicted by the dashed red square. (\textbf{d}) Intensity cross-sections highlight the improvement in contrast provided by dual modulation.
\vspace{5mm}
  }
  \label{fig:block_size_sweep_simulation}
\end{figure}

Next, we showcase how dual modulation performs in simulation, highlighting scenarios where phase-only modulation fails to generate high-contrast images. We evaluate peak signal-to-noise ratio (PSNR) for each image. However, PSNR does not
specifically evaluate contrast, so we also evaluate contrast error (CE), a contrast-specific metric that we describe in Methods.

\subsubsection*{Contrast and high-frequency features}
Using our contrast metric CE, we evaluate our dual modulation approach on 2D image content in Fig.~\ref{fig:block_size_sweep_simulation} for three amplitude SLM pixel sizes:
$\Delta_a = 64$~\textmu m, 480~\textmu m, and 2880~\textmu m (well below, approximately at, and significantly above the condition in Eq.~\eqref{eq:maximum_pixel_size}, respectively).

As predicted, 
phase-only modulation cannot faithfully create content when bright and dark portions of the image are spatially separated, such as the image of vertical bars (Fig.~\ref{fig:block_size_sweep_simulation}a).
In comparison, dual modulation restores contrast, even with large $\Delta_a$, resulting in 27-32 dB increase in PSNR and 95.6-98.9\% decrease in CE.

Figure~\ref{fig:block_size_sweep_simulation}b shows an example of a natural, high-contrast scene, where once again dual modulation significantly outperforms phase-only modulation.
In the case with the largest amplitude pixel size, there is some reduction in PSNR compared to the smaller values of $\Delta_a$, and this difference is also visible in the cross-sections in Fig.~\ref{fig:block_size_sweep_simulation}d.
However, contrast is still visually high in this image, which is accurately captured by the contrast error.

Although the amplitude resolution is low, this does not limit the resolution of our dual modulation approach. %
This is highlighted in the insets in Fig.~\ref{fig:block_size_sweep_simulation}c where the size of the amplitude pixel is indicated in red.
Notice that dual modulation produces high resolution features (e.g. the text on the sign) that are much smaller than the amplitude pixel size.
Dual modulation has the same resolution as phase-only modulation since small features are created with the phase SLM.

\subsubsection*{Focal stacks with dual modulation}

\begin{figure}[!t]
\includegraphics[width=\textwidth]{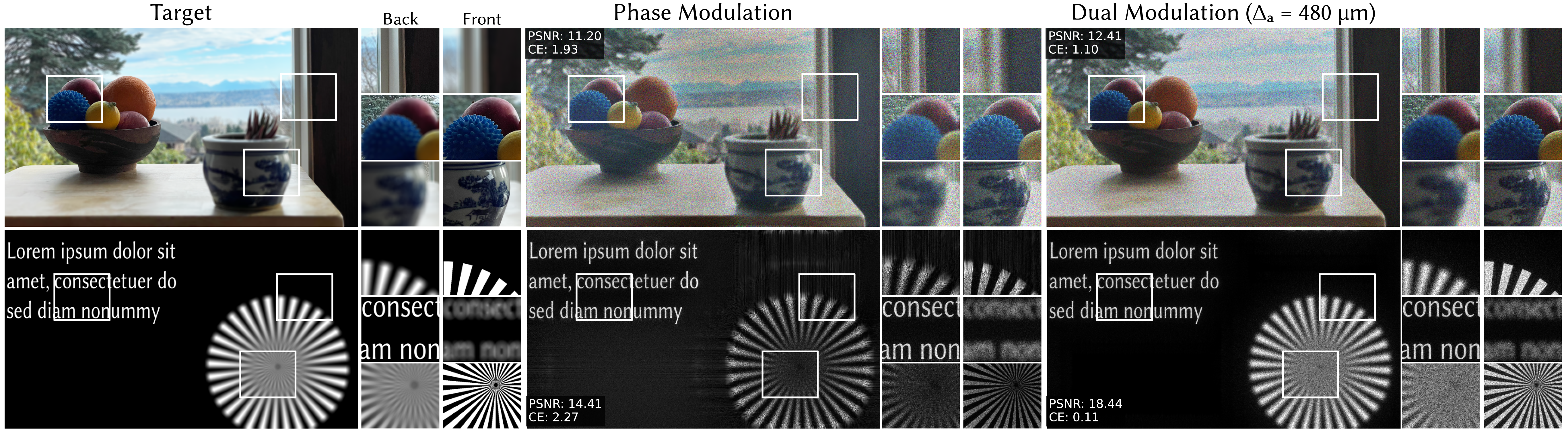}
  \caption{
  \textbf{Focal stacks (simulation).}
  We use dual modulation to generate focal stacks with natural defocus cues, propagating from $20$ to 25~mm with $5$ jointly optimized frames per color channel. Dual modulation, here with $\Delta_a = 480$~\textmu m amplitude pixel size, outperforms phase-only modulation. PSNR and CE calculated over the full focal stack are shown on the left side of each image.}
  \label{fig:3D_system}
\end{figure}

\begin{figure}[!ht]
  \centering
  \includegraphics[width=0.6\textwidth]{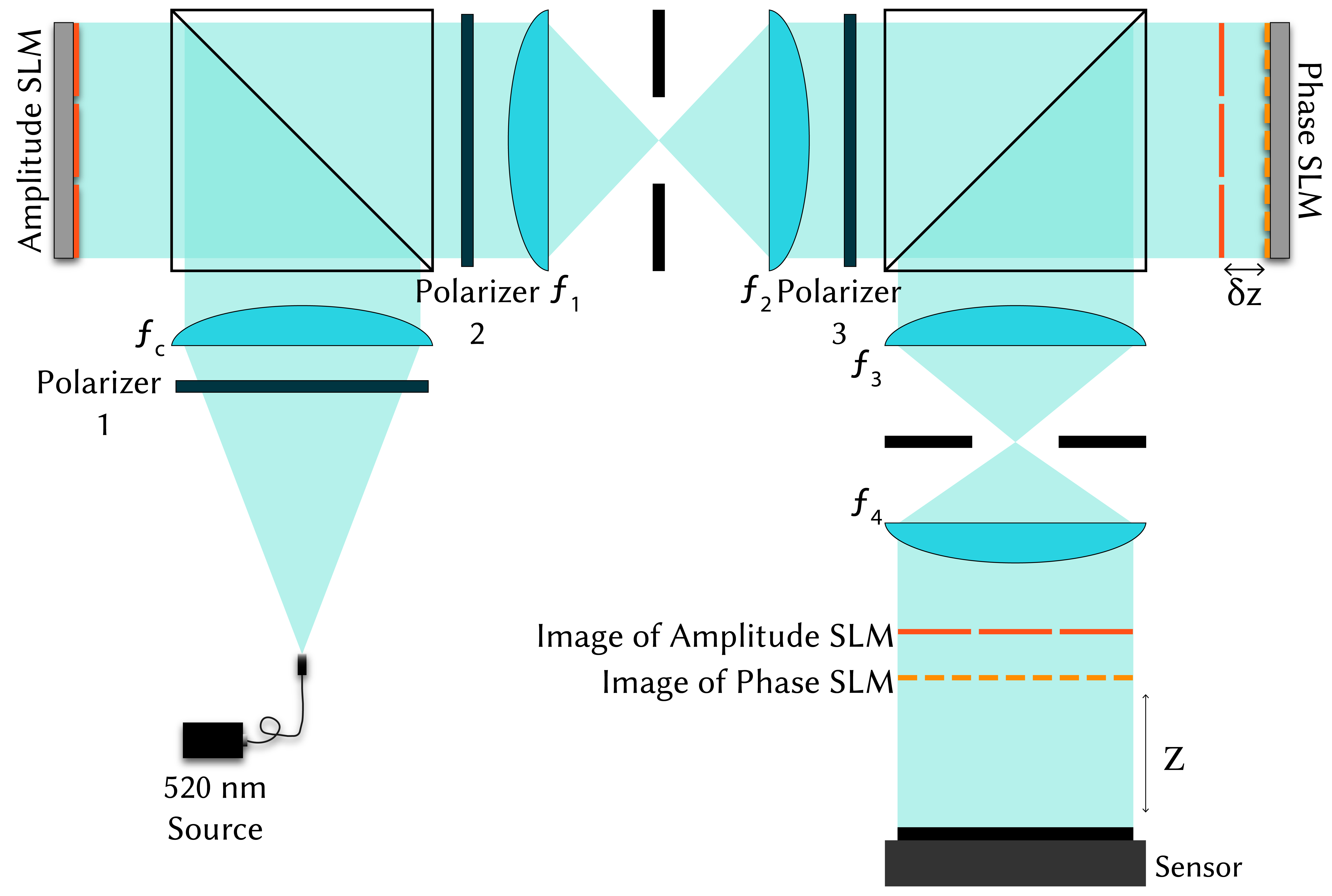}
  \caption{\textbf{Experimental setup schematic.} Our dual modulation prototype uses two reflective SLMs with equal pixel pitch, allowing us to vary the relative pixel size by binning at the amplitude modulator. This requires a 4$f$ system between the SLMs to make room for beamsplitters, but we anticipate our dual modulation approach is compatible with a transmissive amplitude modulator that could enable a compact setup. 
  }
  \label{fig:hardware_setup}
\end{figure}

A key advantage of holographic displays is the ability to generate accommodation cues, and we demonstrate that dual modulation improves contrast for focal stacks as well.
The intuition is that a focal stack maintains the same low-frequency structure at different propagation distances.
Therefore, one low resolution amplitude pattern can provide improved contrast over a range of $z$.
The focal cues themselves are primarily higher resolution features which are created with the phase SLM, just like in a phase-only hologram.

Figure~\ref{fig:3D_system} compares phase-only modulation with dual modulation ($\Delta_a = 480$~\textmu m) for focal stacks with 5 evenly spaced planes from $z = 20$ to 25~mm.
Although both phase-only and dual modulation suffer from speckle noise, which is common for focal stacks without temporal multiplexing~\cite{binaryamplitudeslm22} or other speckle reduction techniques~\cite{gracemultisource23}, dual modulation has visually improved contrast and CE improves by 43-95\% in these examples.

\subsection*{Dual modulation experimental setup and validation}
\label{sec:experimentalsetup}

We experimentally demonstrate dual modulation with a benchtop system, depicted in the schematic in Fig.~\ref{fig:hardware_setup}.
To facilitate experiments with variable $\Delta_a$, we choose to use two identical reflective SLMs (Holoeye Pluto-2.1 VIS-016) where one SLM is set to amplitude mode using crossed polarizers;
the relative resolution of the SLMs is varied by binning pixels on the amplitude modulator.
A 4$f$ system with unit magnification optically relays the SLMs to be $\delta z = 2.4$~mm apart and provides room for beamsplitters to illuminate the reflective modulators.
A second 4$f$ system (mimicking the eyepiece and the eye lens) demagnifies the image onto a sensor (XIMEA MC089MG-SY) at $z = 20$~mm, and both 4$f$ systems include circular apertures in their Fourier planes to filter out SLM higher orders.
The illumination is a fiber-coupled laser diode (FISBA READYBeam Ind 2), and all images are captured at $\lambda = 520$~nm.

Achieving high-quality experimental results requires calibrating for non-idealities in the system such as source non-uniformities, aberrations in the optics, and nonlinear lookup tables of both SLMs.
In addition, dual modulation requires accurate registration between the two SLMs.
We adapt the camera-based calibration approaches of Multisource Holography~\cite{gracemultisource23}, which uses a physics-inspired forward model trained with a dataset of experimentally captured images.
After training, the final images are fine-tuned using the active camera-in-the-loop (CITL) procedure~\cite{peng2020neural}.
Details of the model and training procedure as well as comparisons without CITL can be found in the Supplementary Material.

\begin{figure}[!h]
\includegraphics[width=\textwidth]{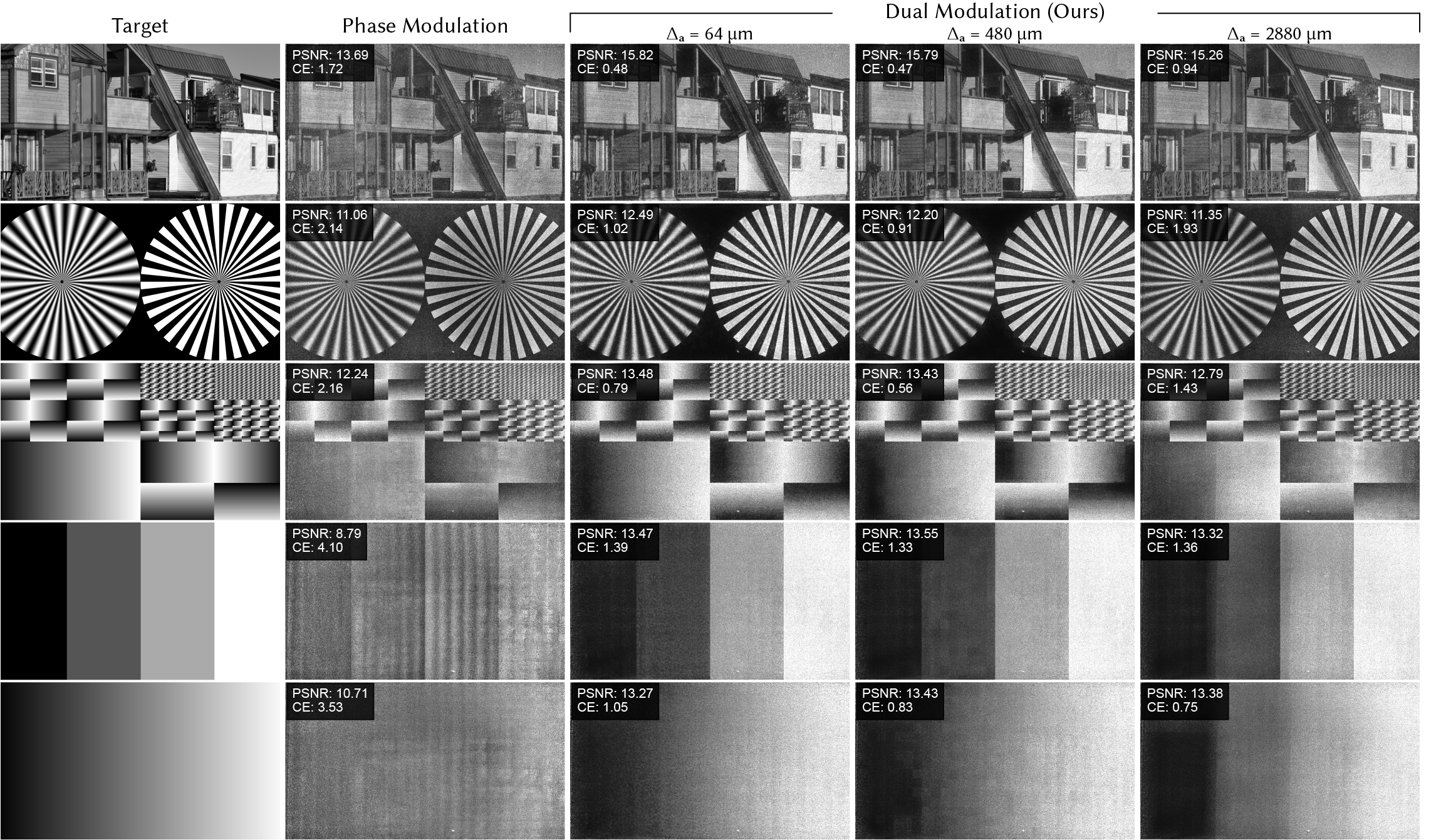}
  \caption{\textbf{Monochrome 2D results (experiment).} We experimentally compare phase-only modulation and dual modulation with three amplitude modulator pixel sizes: $\Delta_a = 64$~\textmu m,  $480$~\textmu m, and $2880$~\textmu m. Phase-only modulation fails to accurately reproduce low-frequency features, for example, the smooth gradient in the bottom row. Dual modulation has clear reduction in contrast error (CE) and improvement in image quality, most clearly seen in low-frequency features. For the lowest resolution dual modulation ($\Delta_a = 2880$~\textmu m), individual amplitude pixels are visible for images with low-frequency features, but less perceptible with high-frequency features, as expected. PSNR and CE are shown in the top left of each experimental capture.
  Houses source image by Madeleine Deaton (CC BY 2.0).
  }
  \label{fig:experimental_grayscale_2D}
\end{figure}

2D monochrome experimental captures from our system are shown in Fig.~\ref{fig:experimental_grayscale_2D}, where we compare three amplitude SLM pixel sizes  ($\Delta_a = 64$~\textmu m, 480~\textmu m, 2880~\textmu m). Phase-only results are captured by holding the transmission of the amplitude SLM constant. 
Although 2D holograms with phase modulation alone are generally high quality, the lack of image contrast is clear when compared to the target or our dual modulation approach. %

Dual modulation restores contrast and image quality, providing up to 1.24-4.76 dB increase in PSNR and 57.5-78.8\% decrease in CE. Notice that the visual image quality and metrics are very similar when $\Delta_a = 64$~\textmu m and $\Delta_a = 480$~\textmu m since both are within the recommended bound of Eq.~\eqref{eq:maximum_pixel_size}. In both cases there are no visible edge effects at the boundaries of amplitude pixels, reinforcing that the larger pixel sizes we propose are sufficient for high contrast images. When the amplitude pixel size does not meet the condition in Eq.~\eqref{eq:maximum_pixel_size} ($\Delta_a = 2880$~\textmu m), the performance is content-dependent. The system performs well when there is high-frequency content in the image (for example, the house on the top row), but struggles with low frequency content like the gradient on the bottom row where the discrete structure of the amplitude pixels is clearly visible.

\begin{figure}[!t]
\includegraphics[width=\textwidth]{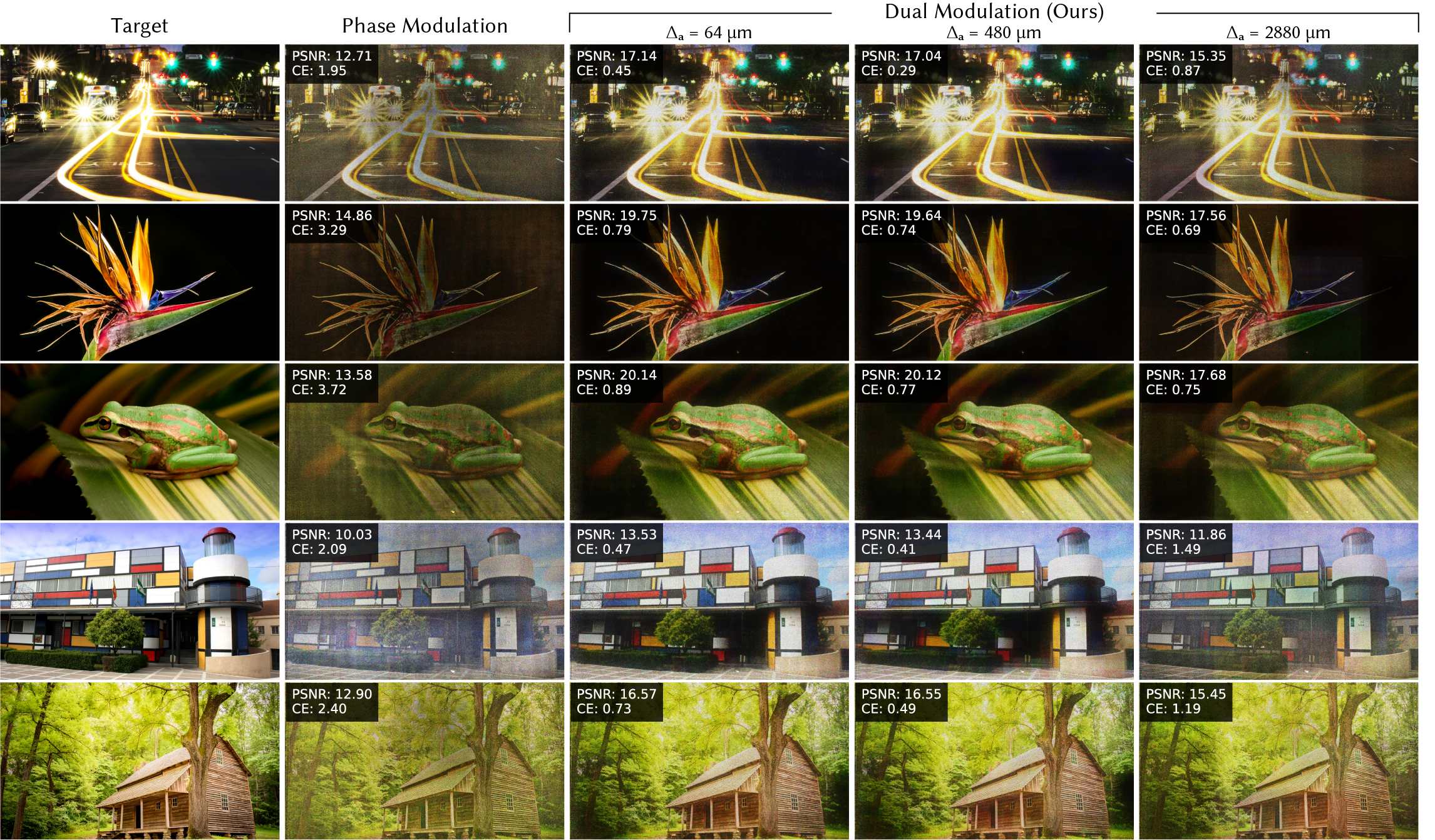}
  \caption{\textbf{Color 2D results (experiment).} To visualize the impact of dual modulation on color images, we experimentally capture pseudocolor images comparing phase-only modulation and dual modulation with three amplitude modulator pixel sizes: $\Delta_a = 64$~\textmu m, $480$~\textmu m, and $2880$~\textmu m. Improved contrast and color fidelity are clear for dual modulation, particularly for dark backgrounds, bright lights, and colorful regions. For the largest amplitude modulator pixel size ($\Delta_a = 2880$~\textmu m), which does not meet the requirement in Eq.~\eqref{eq:maximum_pixel_size}, individual amplitude pixels are sometimes visible; however, quality is similar for both smaller pixel sizes ($\Delta_a = 64$~\textmu m, $480$~\textmu m), validating that there is minimal degradation with larger amplitude pixels as long as Eq.~\eqref{eq:maximum_pixel_size} is met. PSNR and CE are shown in the top left of each experimental capture. Flower source image by Paul Longinidis (CC BY 2.0).
}
  \label{fig:experimental_color_2D}
\end{figure}

To visualize the performance of dual modulation for color images, we capture pseudocolor results, shown in Figure~\ref{fig:experimental_color_2D}. Once again, dual modulation improves image contrast, providing up to 3.50-6.56 dB increase in PSNR and 79.0-85.1\% decrease in CE. Visually, color fidelity is improved as well. As with the monochrome results, when $\Delta_a = 2880$~\textmu m, the images with lower frequency content have visible banding from the low resolution amplitude pixels, but when $\Delta_a$ meets the condition in Eq.~\eqref{eq:maximum_pixel_size}, banding is imperceptible. Since there is minimal improvement when $\Delta_a$ is further reduced, we recommend using pixels near the size proposed in Eq.~\eqref{eq:maximum_pixel_size} since these large pixels can enable high fill factor in transmissive SLMs and save bandwidth without sacrificing image quality.

%% file: current_nat_sci_rep/3_discussion.tex
In summary, we introduce a practical dual modulation approach, combining a phase-only SLM with a low resolution amplitude SLM for improved contrast at short propagation distances $\sim$ 20~mm. We analyze the relationship between the amplitude SLM pixel size and the phase SLM pixel size, and propose a recommended ratio based on diffraction theory. With our approach, we demonstrate up to 31 dB improvement in PSNR in simulation and 6.5 dB experimentally compared to phase-only modulation. Finally, we validate our concept with a benchtop experimental prototype, including a calibration scheme for our dual modulators.

Although there are no SLMs that directly modulate both phase and amplitude, additional optics can be used to combine neighboring pixels to achieve complex modulation from a single display panel. Single-sideband filtering~\cite{singlesideband68, generalizedssb19, binaryamplitudeslm22} enables full complex control from an amplitude SLM but requires access to the Fourier plane via a 4$f$ system, which is not compatible with compact HMDs. Another approach, the double phase amplitude coding method (DPAC), combines pairs of phase-only pixels to produce one complex value. Physically combing the pixel pairs can be done with gratings in the Fourier plane~\cite{binarygratingoptimal12, hmdgrating17, polarizedgrating23}, birefringent materials~\cite{birefringent12}, or arranging pixels in a checkerboard pattern and filtering with a circular aperture~\cite{dpacmaimone17, dpacshi21}. Like single-sideband filtering, these strategies all require a bulky 4$f$ system to access the Fourier plane. In addition, they inherently reduce étendue, which is already limited in a holographic display, by at least a factor of two. Our proposed dual modulation approach requires a second SLM but maintains étendue and can be compatible with a compact architecture without a 4$f$ system, as discussed below and in the Supplementary Material.

Our approach is more similar to prior work using multiple modulators. For incoherent displays, cascaded amplitude modulators improve image contrast, and like our work, the pixel pitches of the modulators do not need to be matched~\cite{damberg20073, seetzen2004high}. Contrast in incoherent displays can be further improved by replacing the rear amplitude modulator with an analog micromirror array~\cite{hoskinson2010light} or phase SLM~\cite{damberg2016high, li2024illumination} that can steer light to form a high dynamic range backlight. However, as we showed in this work, phase-only modulation only creates high contrast content when the propagation distance is large, so these architectures are generally bulky and better suited for projectors than wearable displays.

Multiple modulator approaches have also been explored for coherent holography; these strategies do not sacrifice étendue, but the two SLMs cannot physically occupy the same space, so a 4$f$ relay is needed to optically co-locate the display panels~\cite{phaseandampmod96, twoslmsphasefirst07}. Cascaded modulators without a 4$f$ relay have been explored in simulation~\cite{talbotdualmod17}, but only for the case where the two SLMs had identical pixel pitch. Other configurations include placing one SLM at the image plane and one at the Fourier plane~\cite{halfandhalfslm08}, placing two phase SLMs with binary phase gratings far apart~\cite{twophaseprojector12}, or placing the SLMs in two arms of an interferometer~\cite{dualampslm16, choi2021Michelson}. However, these configurations also sacrifice form factor, and interferometer-based approaches are sensitive to sub-wavelength changes in alignment. 

In our approach, we allow for a small air gap between the two SLMs, which paves the way towards a compact configuration without optical relays. In addition, we generalize the idea of complex modulation by allowing the pixel pitch to vary, and we show that large amplitude pixel sizes are effective at improving contrast. These large pixel sizes are compatible with transmissive SLMs, opening additional paths towards compact architectures.

For example, one proposal for a compact architecture is based on the design proposed by Kim et al.~\cite{bunnyears22}, where a thin waveguide illuminates a reflective phase SLM. This design can support a larger field of view compared to compact systems where the hologram is formed inside the waveguide~\cite{jang2024waveguide, gopakumar2024full}. By adding a transmissive amplitude SLM on the side of the waveguide closer to the eye, light interacts with each SLM once (first phase then amplitude) in a thin stack, enabling a form factor compatible with near-eye displays, as visualized and further discussed in the Supplementary Material (Fig. S3). Although this proposed compact architecture does not have room for higher-order filtering with a 4$f$ system, we show in the Supplementary Material that higher orders from the transmissive amplitude SLM have negligible effect on the image quality when the amplitude SLM pixel pitch is sufficiently high, as proposed in our dual modulation framework. However, additional calibration of sub-pixel structures in the transmissive modulator may be necessary, and higher orders between the two SLMs may still need to be modeled computationally \cite{manuhot21}, filtered with a compact volume grating~\cite{compactgratingchangwon19}, or mitigated using the Talbot effect~\cite{talbotdualmod17}.

Our method uses iterative optimization, and it takes minutes to compute a hologram.
We note that true complex SLMs can produce holograms without iterative methods, enabling real-time rendering, which is an advantage that does not extend to our dual modulation approach.
However, many state-of-the-art holographic displays use iterative approaches (for example, when jointly optimizing temporally multiplexed frames for reduced speckle) and would still require iteration even with a complex modulator.
Approximating the output of iterative algorithms in a computationally efficient way is an important research area. There have already been several examples of this using neural networks~\cite{peng2020neural, dpacshi21}, although only with smooth phase holograms.

Our approach involves attenuating incident light, similar to traditional LCD displays. Although our light efficiency is less than phase-only modulation, our light efficiency is better than a traditional LCD display across various images (Supplementary Material Table S1), with up to 62.78\% better efficiency.
Incorporating a light-efficiency loss~\cite{gordonbrightloss23} could further improve efficiency. %

Our dual modulation approach uses a low resolution amplitude modulator to demonstrate improved contrast for holographic displays at short propagation distances. Low resolution dual modulation serves as an underlying principle that can inform designs for high-contrast displays with compact form factor. We believe that this approach paves the way towards a compact SLM design for high-contrast holography, particularly for head-mounted applications including virtual and augmented reality.

%% file: current_nat_sci_rep/4_materials_and_methods.tex
\subsection*{Experimental setup details}

Our benchtop experimental prototype for dual modulation (Fig.~\ref{fig:hardware_setup}) uses a fiber-coupled laser diode (FISBA READYBeam Ind 2, $\lambda = 520$~nm) illumination source and a collimating lens ($f_c = 500$~mm, Thorlabs AC508-500-A-ML). This prototype uses two identical SLMs, which are Holoeye Pluto-2.1 phase-only reflective liquid crystal on silicon (LCOS) SLMs with resolution of 1920 $\times$ 1080 pixels, pixel pitch of 8~\textmu m, and a bit depth of 8 bits. The first SLM is set to amplitude mode with two crossed polarizers (Polarizer 1 and Polarizer 2, Thorlabs LPVISE200-A) and its resolution $\Delta_a$ is varied by binning pixels. Two Pellicle beamsplitters (Thorlabs BP245B1) are used to illuminate the reflective modulators.

A 4$f$ system with unit magnification ($f_1 = f_2 = 200$~mm, Thorlabs AC508-200-A-ML) relays the amplitude SLM behind the phase SLM such that the two SLMs are a short spacing $\delta z = 2.4$~mm apart. This spacing is a small, non-zero distance chosen heuristically to reduce form factor. Analysis of the design space of SLM spacing and ordering indicates that SLM spacing has minimal impact on image quality (Supplementary Material Fig. S2). Another polarizer (Polarizer 3, Thorlabs LPVISE200-A) converts the incident light to linear polarization for modulation by the phase SLM. 
A demagnifying 4$f$ system ($f_3 = 200$~mm, Thorlabs AC-508-200-A-ML, $f_4 = 150$~mm, Thorlabs AC-508-150-A-ML) relays the hologram to the image plane for capture at $z=20$~mm on a XIMEA MC089MG-SY sensor. There are circular apertures in the Fourier planes of both of the 4$f$ systems to optically filter out SLM higher orders. The apertures do not have zero-frequency (DC) blocks. 

All images are captured at $\lambda = 520$~nm. Color results are captured sequentially and visualized in pseudocolor. Relative color channel intensities are adjusted. Final captured images include active CITL fine-tuning. The Supplementary Material contains additional details about the experimental setup, physics-inspired forward model and training procedure used to calibrate for system non-idealities, and comparisons without CITL.

\subsection*{Hologram generation and software implementation}

Our image formation models use the angular spectrum method (ASM) propagation operator $\mathcal{A}_z$,
\begin{align}
\mathcal{A}_z\left\{\mathbf{G}(\vec{x}) \right\} &=
\mathcal{F}^{-1} \bigl\{ \mathcal{F} \left\{ \mathbf{G}(\vec{x})\right\} \odot \mathcal{H}_z(\vec{u}) \bigr\}, 
\label{eq:ASM_propagation} \\
\mathcal{H}_z(\vec{u}) &= 
\begin{cases}
\exp \left( j \frac{2 \pi}{\lambda} z
\sqrt{1 - \| \lambda \vec{u} \|^2 } \right), & \text{ if } \| \vec{u} \| < \frac{1}{\lambda}, \\
0, & \text{ otherwise, }
\end{cases}
\label{eq:ASM_kernel}
\end{align}
where $\mathcal{F}\left\{\cdot\right\}$ is the Fourier transform operator, $\vec{u}$ is the 2D frequency coordinate, and $\lambda$ is the illumination wavelength~\cite{goodman2005fourieroptics}.

To generate a hologram using one SLM, we use gradient descent with Adam~\cite{kingma2014adam} to determine the optimal SLM modulation $\mathbf{m}$ that produces the desired intensities at the target image planes based on the image formation model in Eq.~\eqref{eq:phase_modulation_model}:
\begin{equation}
    \mathbf{m} = \operatorname*{argmin}_\mathbf{m} \sum_z \| \mathbf{I}_z(\vec{x}) - \hat{\mathbf{I}}_z(\vec{x}) \|_2^2.
    \label{eq:phase_only_optimization}
\end{equation}
Note that the intensity of the displayed image, $\mathbf{I}_z(\cdot)$ can be scaled arbitrarily by changing the intensity of the incident field, $\mathbf{G}_0(\cdot)$. To remove this factor from the optimization, we normalize both $\mathbf{I}_z(\cdot)$ and $\hat{\mathbf{I}}_z(\cdot)$ by their respective means, which ensures the target and displayed images have equal energy over the image plane. 

In addition, we initialize $\mathbf{m}(\vec{x})$ with uniform random phase to generate ``random phase'' holograms which scatter light evenly over all possible diffraction angles. Although ``smooth phase'' holograms (generated by initializing $\mathbf{m}(\vec{x})$ with constant phase) tend to have less speckle, especially for focal stacks, they are less effective at creating natural defocus blur~\cite{randomphase21} and driving the human accommodation response~\cite{randomisbetter22}. The importance of creating random phase holograms is further discussed in~\cite{chakravarthula2022pupil, schiffers2023stochastic, gracemultisource23, shi2024ergonomic, randomphasereview24}.

The algorithm for generating a hologram for dual modulation is a dual variable optimization based on the image formation model in Eq.~\eqref{eq:dual_modulation_model}:
\begin{equation}
\begin{aligned}
    &\mathbf{m_a}, \mathbf{m_p} = \operatorname*{argmin}_{\mathbf{m_a}, \mathbf{m_p}} \sum_z \| \mathbf{I}_z(\vec{x}) - \hat{\mathbf{I}}_z(\vec{x}) \|_2^2 \\
    \text{s.t.} \quad \mathbf{m_a}(\vec{x}) = &\text{ const.} \quad \text{while} \quad \vec{x}_i \leq \vec{x} < \vec{x}_i + \vec{\Delta}_a \quad \text{for each} \ i,
    \label{eq:dual_optimization}
\end{aligned}
\end{equation}
where $\vec{x}_i$ is the spatial coordinate at the corner of each amplitude pixel, $i$; $\vec{\Delta}_a = [\Delta_a, \Delta_a]^T$; and the inequalities are element-wise over the two spatial coordinates. Together, this constraint restricts $\mathbf{m_a}(\vec{x})$ to have a constant value within each pixel of size $\Delta_a$, allowing the pattern to be shown on a low resolution SLM.

All algorithms are implemented in PyTorch~\cite{pytorch}. 
Images are optimized by solving Eq.~\eqref{eq:phase_only_optimization}, for phase-only modulation, and Eq.~\eqref{eq:dual_optimization}, for dual modulation, on an NVIDIA A6000 GPU with $\Delta_p = 8$~\textmu m, $z = 20$~mm, $\delta z = 2.4$~mm, and the amplitude SLM located before the phase SLM.
Monochrome simulations are conducted in green ($\lambda = 520$~nm), and color simulations are conducted in red-green-blue ($\lambda = 638$~nm, 520~nm, 450~nm).
The phase SLM is initialized with uniform random phase and the amplitude SLM with constant amplitude.

For color images, PSNR and CE are averaged across the three color channels.
For focal stacks, the metrics are averaged across all depth planes and the three color channels. 
Increase in PSNR and decrease in $\operatorname{CE}$ reported are based on the dual modulation amplitude pixel size that produced the best metric values for each image.

\subsection*{Contrast error (CE) metric}

As PSNR does not specifically evaluate contrast, we developed a contrast-specific metric, contrast error (CE), which is used throughout the text. Our metric builds upon a common contrast metric, the Michelson contrast, 
\begin{equation}
    \mathcal{C}(\mathbf{I}) = \frac{\mathbf{I}_{\max} - \mathbf{I}_{\min}}
    {\mathbf{I}_{\max} + \mathbf{I}_{\min}}
    \label{eq:michelson_contrast},
\end{equation}
which compares the highest and lowest intensities across an image. Holographic images are prone to speckle noise due to coherent illumination, which can give a false boost to the lightest and darkest intensity values, resulting in artificially high contrast.
In addition, as described in our analysis, %
we expect contrast in holograms to vary based on spatial frequency, which we'd like our metric to capture.

Therefore, we first represent the image as a Gaussian pyramid by recursively low-pass filtering and downsampling the image to form a multi-level representation~\cite{gaussianpyramid83}.
At each Gaussian pyramid level, we calculate the resulting image's Michelson contrast for both the target and displayed images.
Finally, we sum the absolute error over all pyramid levels.
We call the resulting value ``contrast error'' (CE), where smaller values indicate a more faithful representation of the target.
See the Supplementary Material (Fig. S1) for further implementation details.

$\operatorname{CE}$ percentage improvement reported in the text is computed as
\begin{equation}
    \frac{\operatorname{CE}_p - \operatorname{CE}_d}{\operatorname{CE}_p},
\end{equation}
where $\operatorname{CE}_p$ is the contrast error for phase-only modulation and $\operatorname{CE}_d$ is the contrast error for dual modulation, evaluated on the same image.

%% file: current_nat_sci_rep/6_supplement_text.tex
\section{Contrast error metric}
Peak signal-to-noise ratio (PSNR) captures image quality, but does not specifically evaluate contrast. 
We developed a multi-level contrast metric to evaluate contrast performance.

As described in the main text, a common contrast metric is the Michelson contrast, 
\begin{equation}
    \mathcal{C}(\mathbf{I}) = \frac{\mathbf{I}_{\max} - \mathbf{I}_{\min}}
    {\mathbf{I}_{\max} + \mathbf{I}_{\min}}
    \label{eq:michelson_contrast_supp},
\end{equation}
which compares the highest and lowest intensities across an image. 
When the Michelson contrast is evaluated on a holographic image, speckle noise can give a false boost to the contrast value. 
Furthermore, evaluating the Michelson contrast on a full image does not capture variations in contrast based on spatial frequency. 

In order to thoroughly evaluate contrast performance at multiple spatial frequencies, we represent the image as a Gaussian pyramid~\cite{gaussianpyramid83}.
Each increasing Gaussian pyramid level captures lower-frequency image properties from low-pass filtering and downsampling operations.
We compute Gaussian pyramids for both the target and the displayed hologram. 
At each Gaussian pyramid level, we compute the Michelson contrast for the target and the displayed hologram, and calculate the absolute error between target and hologram contrast. 
Finally, we sum the absolute error over all pyramid levels. 
We call this multi-level contrast metric ``contrast error'' (CE). An image with accurate contrast aims to minimize this quantity.

The equation to compute CE is
 \begin{equation}
\operatorname{CE}\left(\mathbf{I}, \hat{\mathbf{I}}\right) = 
\sum_{g} | \mathcal{C} ( \mathcal{G}(\mathbf{I})) - \mathcal{C}( \mathcal{G}(\hat{\mathbf{I}})) |, 
 \label{eq:contrast_error_metric}
 \end{equation}
 in which the Michelson Contrast $\mathcal{C}(\cdot)$ is calculated on each Gaussian pyramid level $g$ after applying the Gaussian pyramid $\mathcal{G}(\cdot)$ operator (low-pass filtering and downsampling). Here, $\mathbf{I}$ represents the target image and $\hat{\mathbf{I}}$ represents the displayed hologram.  Better contrast images have lower $\operatorname{CE}$.

\begin{figure}
  \includegraphics[width=\textwidth]{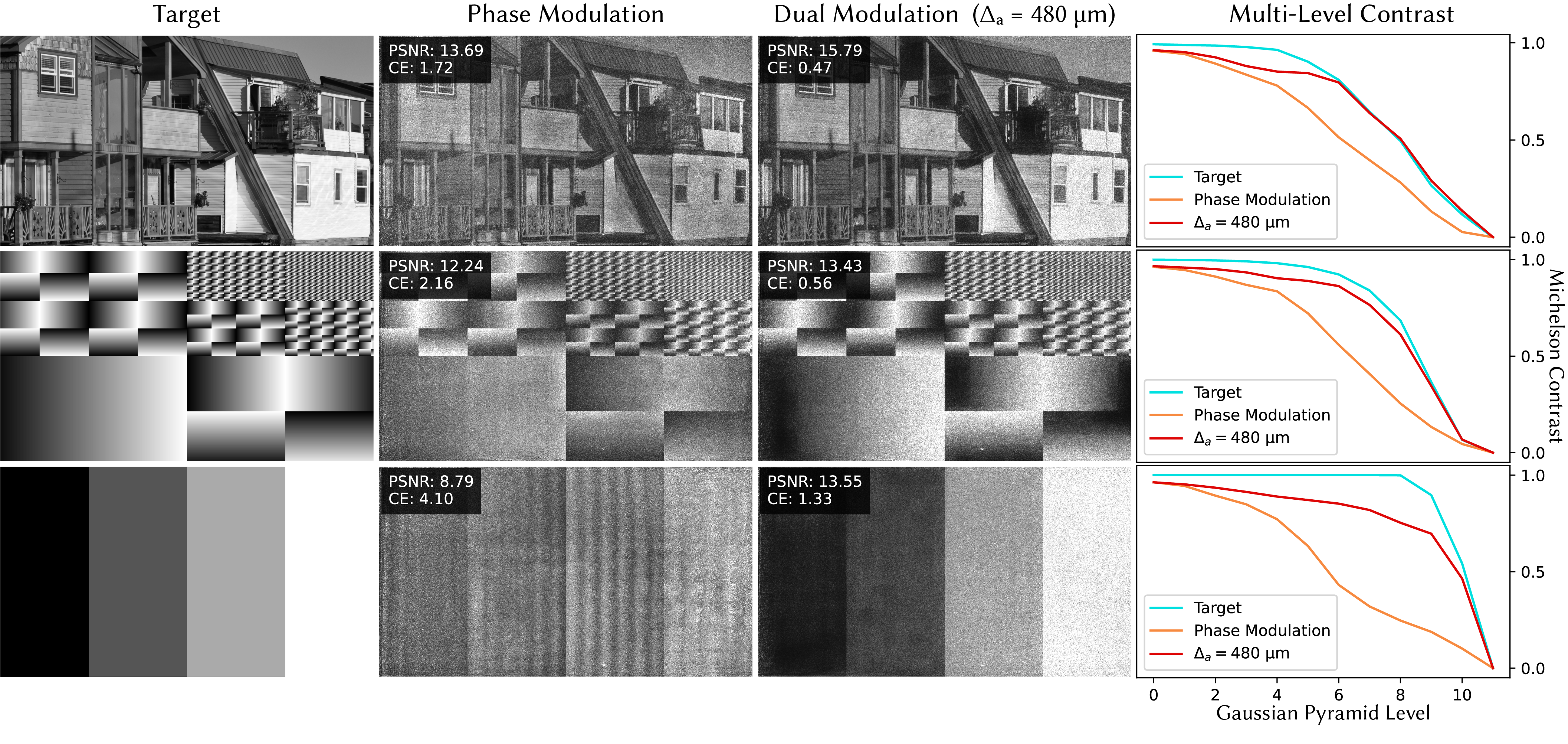}
  \caption{\textbf{Contrast error (CE) metric.} Experimentally captured images and their corresponding multi-level contrast plots. To form the contrast plots, the Michelson contrast is computed at all Gaussian pyramid levels for each experimental capture and target. Gaussian pyramid level 0 corresponds to the full image and increasing pyramid levels correspond to increasing downsampling. The absolute error between the contrast for the target and the captured image is summed across Gaussian pyramid levels to produce the CE value for each image.
  PSNR and CE are shown in the top left of each image. Houses source image by Madeleine Deaton (CC BY 2.0).
}
  \label{fig:contrast_error}
\end{figure}

Figure~\ref{fig:contrast_error}  shows experimentally captured images with both phase-only modulation and dual modulation ($\Delta_a = 480$~\textmu m) and their corresponding multi-level contrast plots, which contain the Michelson contrast at each Gaussian pyramid level.
Insets in the top left of each experimental capture show PSNR and CE.
Pyramid level 0 corresponds to the full image and increasing pyramid levels correspond to further downsampled images. 
To compute CE with Eq.~\eqref{eq:contrast_error_metric}, we sum the absolute error between the target and the hologram contrast across Gaussian pyramid levels.
\section{Analysis: modulator order and spacing}
\label{sec:modulator_config_analysis}
\begin{figure}[!t]
  \centering \includegraphics[width=0.45\textwidth]{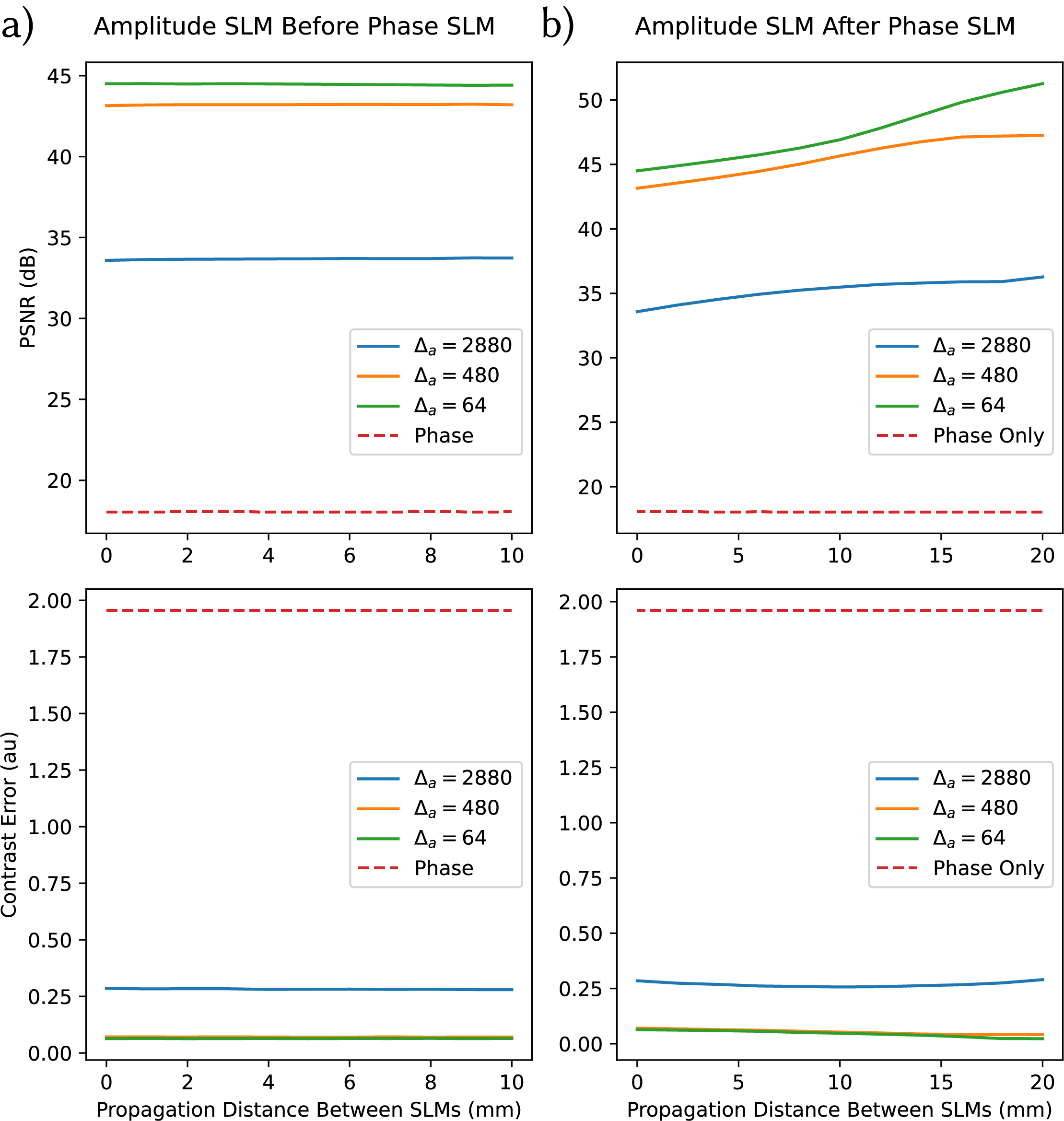}
  \caption{\textbf{Dual modulation system configuration analysis.} The ordering and spacing between the amplitude and phase spatial light modulator (SLM) affects PSNR and Contrast Error (CE). (\textbf{a}) With the amplitude SLM placed before the phase SLM, the spacing between the modulators does not affect metrics. (\textbf{b}) With the amplitude SLM placed after the phase SLM, performance depends on the spacing between the modulators. Regardless of the configuration, dual modulation provides significant improvement compared to phase only modulation. Metrics are averaged over the set of grayscale images used in the main text.}
  \label{fig:system_configuration}
\end{figure}
In the main text, we analyzed the effect of amplitude SLM pixel size on contrast and image quality. Other available design parameters are the ordering of the amplitude and phase SLMs and the spacing between the two SLMs. Here we study these parameters for three amplitude SLM pixel sizes ($\Delta_a = 64$~\textmu m, $480$~\textmu m, and $2880$~\textmu m). The phase SLM  pixel size is fixed at $\Delta_p = 8$~\textmu m in this analysis.

First we consider the case where the amplitude SLM is placed before the phase SLM, and results are shown in Fig.~\ref{fig:system_configuration}a. In this configuration, the propagation distance after the phase SLM is fixed at 20~mm and the distance between the amplitude and phase SLM is swept between $0$ and 10~mm. Metrics (PSNR and CE) are averaged over the set of grayscale images used in the main text. We see that when the amplitude SLM is first, the hologram quality does not depend on the distance between the two SLMs. This is the configuration that we demonstrated in experiment in the main text, with $\delta_z = 2.4$~mm.

Next, we consider the case where the phase SLM is placed before the amplitude SLM, and results are shown in Fig.~\ref{fig:system_configuration}b. Here, the total propagation distance, measured from the phase SLM to the image plane, is fixed at 20~mm. Then, the position of the amplitude SLM is swept across possible positions between the phase SLM and the image plane ($\delta_z = 0$ to 20~mm). In this configuration, improvement from dual modulation depends on the propagation distances, with better performance as the amplitude modulator is closer to the image plane. However, there is clear improvement at all distances tested compared to phase only. For short propagation distances between the SLMs ($\delta_z < {\sim} 5$~mm), the performance is similar regardless of which SLM is placed first.

\subsection{Compact architectures}

\begin{figure}[!t]
    \centering \includegraphics[width=0.9\textwidth]{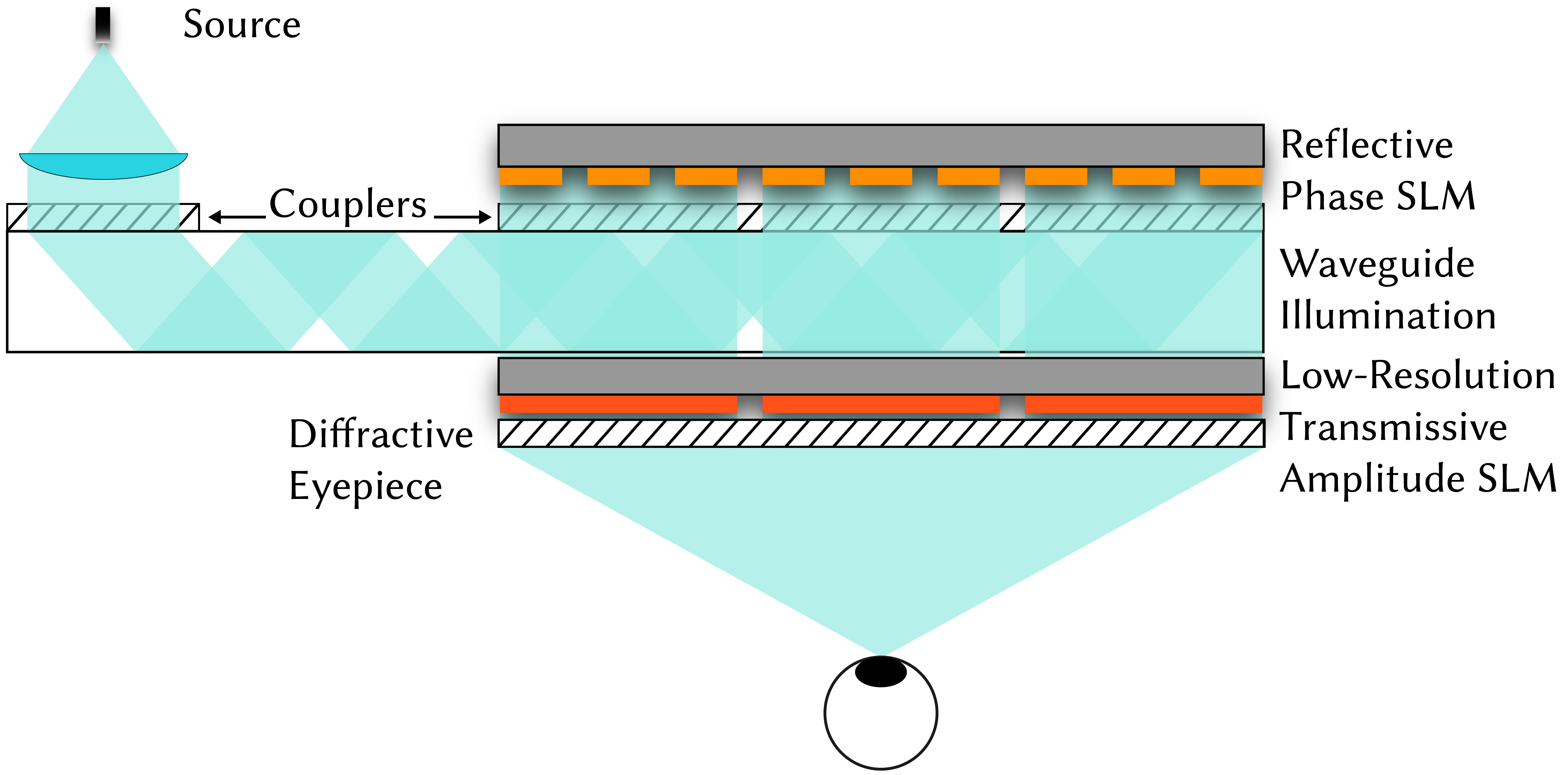}
    \caption{\textbf{Compact architecture example.} An example of a compact architecture for dual modulation places a reflective phase spatial light modulator (SLM) and a transmissive amplitude SLM on two sides of a thin waveguide, which couples the illumination from the source to the phase SLM. In this configuration light interacts first with the phase SLM and then with the amplitude SLM, in a thin form factor suitable for near-eye displays.}
    \label{fig:example_compact_diagram}
\end{figure}

In the main text, we focused on the configuration with the amplitude SLM placed first, which was demonstrated with a benchtop experimental prototype. The alternative configuration that places the phase SLM first is well suited for compact architectures, with form factor compatible with near-eye displays. An example of such a compact architecture is visualized in Fig.~\ref{fig:example_compact_diagram}. This design is based on the one proposed by Kim et al.~\cite{bunnyears22}, where a thin waveguide couples the illumination from a source to a reflective phase SLM. For dual modulation, a transmissive amplitude SLM can be added on the opposite side of the waveguide, closer to the eye. In this thin optical stack, light interacts with each SLM once, first with the phase SLM and then with the amplitude SLM. Our dual modulation approach may also be applied to other viable options for compact architectures, such as those involving multiple interactions with the amplitude SLM.

\section{Analysis: higher-order effects from amplitude modulator}
\label{supp:higher_order_analysis}

\begin{figure}[!t]
  \includegraphics[width=\textwidth]{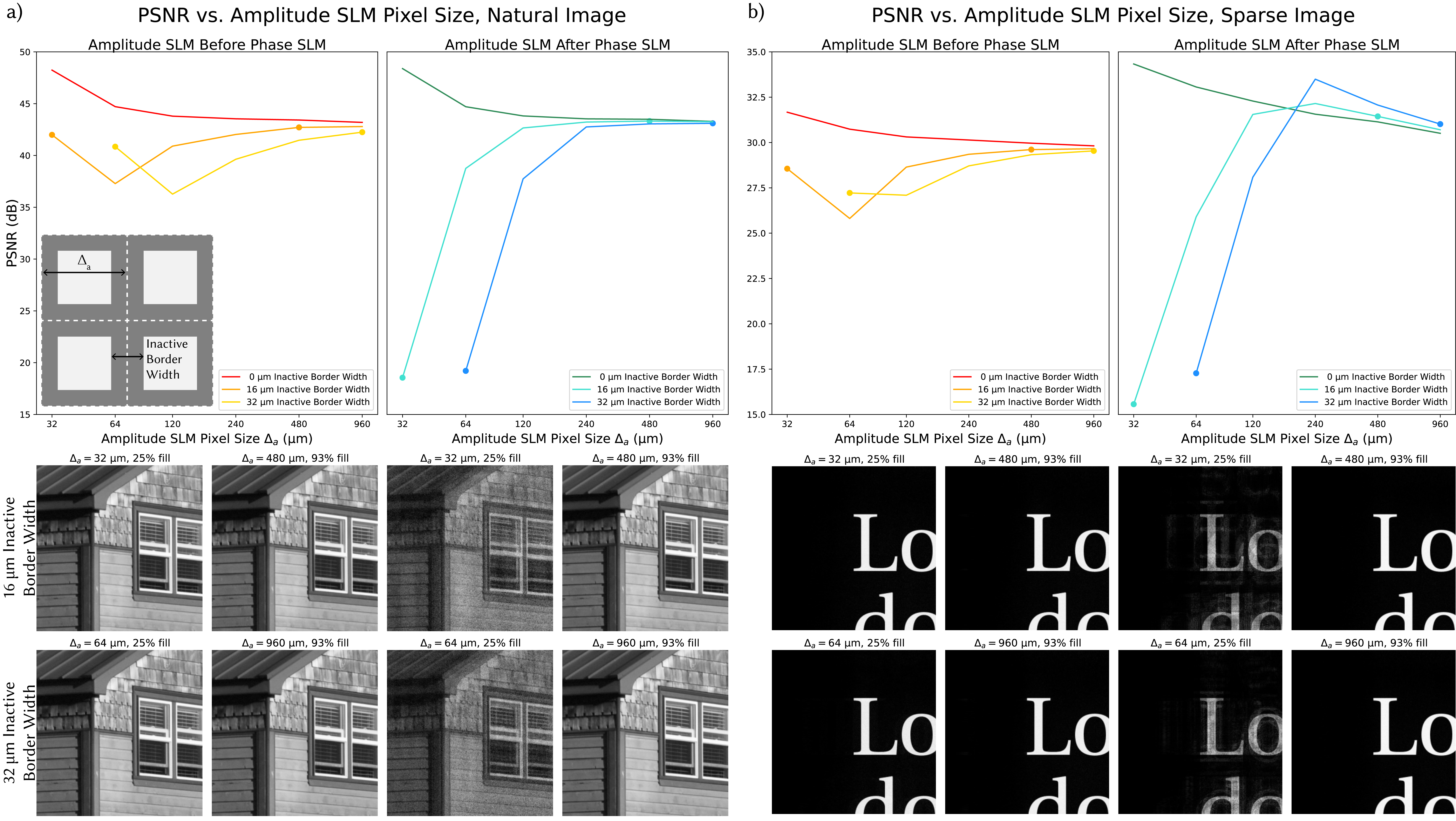}
  \caption{\textbf{Amplitude spatial light modulator (SLM) fill factor.} We analyze the impact of higher order effects from the amplitude SLM by simulating an inactive, opaque border around each amplitude pixel (depicted in the upper left inset). We compare several inactive border widths over a range of pixel sizes for both (\textbf{a}) a natural image of a house and (\textbf{b}) a sparse image of text. Note that border width of 0~\textmu m corresponds to perfect fill factor, which will minimize higher orders. When the amplitude SLM is first, higher order effects can mostly be compensated for by the phase SLM, and there is minimal degradation from the incomplete fill factor. When the amplitude SLM is second, higher orders cause doubling in the image when the pixel size is small. However, for the large pixel pitches proposed in this work (${\sim} 480$~\textmu m), peak signal-to-noise ratio (PSNR) is high, and there is visually no doubling, even with a significant inactive border around each pixel. Houses source image by Madeline Deaton (CC BY 2.0).
}
  \label{fig:higher_orders}
\end{figure}

Our experimental setup uses multiple 4$f$ systems, which allow for filtering of the higher orders that arise from the pixel structure of the SLMs. However, as described in the main text, 4$f$ systems are not compatible with a compact system, so it's advantageous if higher orders can be accounted for computationally instead of physically filtered. In particular, we propose using a transmissive amplitude modulator. Transmissive modulators generally have stronger higher-order effects due to their lower fill factor (compared to reflective modulators). Here, we show in simulation that given the large pixel pitches in dual modulation, fill factor is high and higher orders have negligible impact on image quality if the sub-pixel structure of the transmissive pixels is known.

Specifically, we simulate the case of reduced fill factor at the amplitude modulator. In a transmissive SLM, the electronics block light at each pixel, so we model the sub-pixel structure as an opaque, inactive border of fixed width around each pixel (see diagram in Fig.~\ref{fig:higher_orders}a). We assume that the inactive region is known in advance, and we take it into account during each iteration of optimization by multiplying the amplitude SLM pattern by a binary mask representing the inactive region. We simulate a range of amplitude pixel sizes from 32~\textmu m to 960~\textmu m while holding the inactive border width constant since the size of the electronics is independent of the pixel pitch. We consider three different border widths: 32~\textmu m, 16~\textmu m, and 0~\textmu m, which corresponds to the case of perfect fill factor. In total, the simulation covers fill factors ranging from 25\% to 100\%.

We simulate both the configuration with the amplitude SLM first and with the phase SLM first. When the amplitude SLM is first, we simulate $\delta_z = 2.4$~mm between the amplitude SLM and phase SLM, followed by $z = 20$~mm propagation to the detector plane. 
When the phase SLM is first, we simulate $\delta_z = 4$~mm between the phase SLM and amplitude SLM with $z = 16$~mm propagation after the amplitude SLM to the detector plane. Note that the case with the amplitude modulator second could enable a compact architecture, as described in the main text. The simulated illumination is $\lambda = 520$~nm.

Fig.~\ref{fig:higher_orders} shows the simulation results on both  a natural image (a) and a sparse image (b). For the configuration with the amplitude SLM before the phase SLM (Fig.~\ref{fig:higher_orders}, left), higher-order effects are imperceptible across pixel sizes for both the natural and sparse image. There is a slight overall reduction in PSNR compared to the reference (0~\textmu m border width) due to a portion of the incident illumination on the phase SLM being blocked by the amplitude SLM pixel structure. However, we see that PSNR remains high, even for larger inactive border widths (which correspond to smaller fill factor).

For the configuration with the amplitude SLM after the phase SLM (Fig.~\ref{fig:higher_orders}, right), smaller amplitude SLM pixel sizes ($\Delta_a = 32$~\textmu m, $\Delta_a = 64$~\textmu m) have visible higher orders due to the low fill factor. These higher orders are clearly visible as overlapping replicas in the sparse image of text. However, as amplitude SLM pixel size increases, image quality improves, and for large amplitude SLM pixel sizes ($\Delta_a = 480$~\textmu m, $\Delta_a = 960$~\textmu m), there are no visible higher orders and the PSNR is almost the same as the reference. In fact, with the sparse image, the unique combination of image content and pixel border structure can provides a slight boost in PSNR. As a reminder, in the main text we show that pixel pitches as large as  $\Delta_a = 480$~\textmu m are sufficient to generate high quality, high contrast imagery. Even assuming much more conservative border widths (16 - 32~\textmu m) than the minimum suggested by Curatu and Harvey~\cite{curatu2009analysis} (2.8~\textmu m), we see that higher orders have negligible effect at these larger pixel sizes.

\section{Dual modulation light efficiency}

\begin{table}[ht]
\centering

\begin{tabular}{ |>{\centering\arraybackslash}p{3cm}||>{\centering\arraybackslash}p{2.5cm}|>{\centering\arraybackslash}p{2.5cm}|>{\centering\arraybackslash}p{2.5cm}|>{\centering\arraybackslash}p{2.5cm}|}

 \hline
  Image & Transmissive Display & Dual Modulation ($\Delta_a = 64$ \textmu m) & Dual Modulation ($\Delta_a = 480$ \textmu m) & Dual Modulation ($\Delta_a = 2880$ \textmu m\\
 \hline
 Street (Fig. 3)   & 0.669    & 0.500 / 25.23\% & 0.486 / 27.28\%  & 0.654 / 2.09\%\\
 Stripes (Fig. 6)&  0.500  & 0.506 / -1.13\% & 0.405 / 18.97\% & 0.420 / 15.97\%\\
 Ramp (Fig. 6) & 0.500 &     0.454 / 9.35\% & 0.368 / 26.40\% & 0.412 / 17.57\%\\
 Gradient (Fig. 6) & 0.500  &  0.468 / 6.54\% & 0.409 / 18.27\% &  0.453 / 9.37\%\\
 Star (Fig. 6)&   0.609  & 0.432 / 28.96\% & 0.265 / 56.44\% & 0.540 / 11.33\%\\
 House (Fig. 6)& 0.566  & 0.541 / 4.44\% & 0.514 / 9.18\%   & 0.691 / -22.14\%\\
 Text (Fig. S4)& 0.885  &  0.558 / 37.00\% & 0.512 / 42.21\% & 0.330 / 62.78\%\\
 \hline
\end{tabular}
\caption{\textbf{Light attenuation for traditional transmissive display vs. dual modulation.} Average light attenuation for a traditional transmissive display and our dual modulation approach at three amplitude pixel sizes  ($\Delta_a = 64$~\textmu m, 480~\textmu m, 2880~\textmu m). 1 corresponds to full attenuation, 0 corresponds to no attenuation, and lower values correspond to better light efficiency. For each dual modulation pixel size we also include the percentage improvement in light efficiency over the transmissive display. Values are reported for each grayscale image used in this work. Dual modulation is up to 62.78\% more efficient than a traditional transmissive display, and on average 19.34\% more efficient.}
\label{tab:lightefficiency}
\end{table}
As discussed in the main text, our approach involves attenuating incident light, as we combine a phase SLM with an amplitude SLM. A phase-only modulation approach is the most light-efficient, as it simply redirects light without attenuating. In contrast, a traditional LCD display is less light-efficient, as it produces images by attenuating light. Here, we show that our approach lands in between these approaches, as it is not only comparable to a traditional LCD display in light efficiency, but in fact more light-efficient for most image content.

For a grayscale image and corresponding amplitude pattern $a$ normalized to the range $0 - 1$, we define its attenuation as $ 1 - a$, where each pixel of the amplitude pattern is subtracted from 1. The average attenuation, or light attenuation, is taken across all pixels of the attenuation, and a lower light attenuation value, corresponding to higher light efficiency, is desirable.

We quantify the light efficiency of a traditional LCD display by considering a transmissive display where the image is formed solely by an amplitude pattern, where the amplitude pattern is the normalized image. For dual modulation, the light efficiency depends on the amplitude pattern resulting from the joint optimization of the phase and amplitude SLM patterns, also normalized. We can then compare the light efficiency of these two approaches by calculating the light attenuation in each case. 

Table~\ref{tab:lightefficiency} reports the light attenuation for both a traditional transmissive display and our dual modulation approach for the set of grayscale images used throughout this work. For dual modulation, we consider the three amplitude SLM pixel sizes ($\Delta_a = 64$~\textmu m, 480~\textmu m, 2880~\textmu m) used in the main text. We additionally calculate the percent improvement in light efficiency of each dual modulation case relative to the traditional transmissive display. Dual modulation's light efficiency benefits are clear for sparse content (ex. Text, Star), high-frequency content (Street, Ramp), and low-frequency content (Gradient). 
For low-frequency content there is one case (Stripe) where dual modulation with $\Delta_a = 64$ \textmu m is comparable to, but does not provide an improvement over a transmissive display, which we attribute to optimization noise. For the Stripe image, lower dual modulation resolutions maintain improvement in light efficiency. For high-frequency content there is one case (House) where the lowest resolution dual modulation with $\Delta_a = 2880$ \textmu m, which is beyond the recommended bound for full light redirection of image content, is less light-efficient than a transmissive display. Overall, dual modulation is up to 62.78\% more light-efficient and on average 19.34\% more light-efficient than a traditional LCD display. 

\section{Experimental prototype}
\begin{figure}[!t]
  \includegraphics[width=\textwidth]{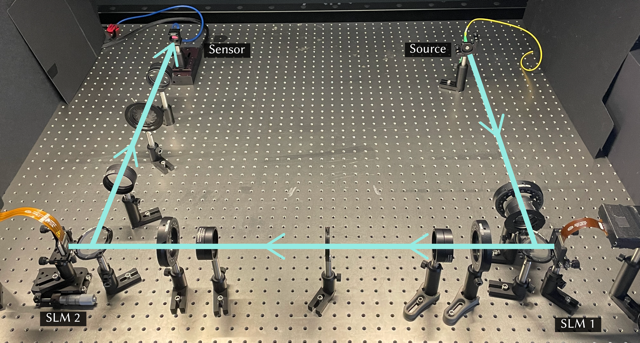}
  \caption{\textbf{Benchtop experimental prototype.} Photograph of benchtop experimental setup with optical path highlighted.}
  \label{fig:setup_picture}
\end{figure}

A photograph of our benchtop experimental prototype is shown in Fig.~\ref{fig:setup_picture}. 
As described in the main text, we use two SLMs. 
A fiber-coupled laser diode ($\lambda = 520$~nm) is the illumination.
Crossed polarizers convert the first SLM to modulate amplitude (SLM1).
A 4$f$ system relays the amplitude SLM to a short spacing $\delta_z = 2.4$~mm behind the phase SLM (SLM2).
Another polarizer converts incident light to linear polarization for modulation by the phase SLM.
A second 4$f$ system relays the SLMs to the sensor plane.
Apertures in the Fourier plane of each 4$f$ system block higher orders.

\section{Calibration details}
\label{sec:calibration}
\subsection{Calibration: dual spatial light modulator (SLM) calibration and modeling}
\label{sec:calibration_modeling}
Achieving high-quality experimental results with a dual modulation system introduces additional challenges with system parameter calibration and SLM alignment and registration.
To address these, we design a forward model for a dual SLM setup, adapting the approaches of Multisource Holography~\cite{gracemultisource23} and Neural Holography~\cite{peng2020neural}. Our system model incorporates several components, including a source model (Sec.~\ref{sec:learnable_source}), propagation model (Sec.~\ref{sec:propagation_model}), phase and amplitude SLM models (Sec.~\ref{sec:phase_amplitude_slm_models}), a dual SLM alignment model (Sec.~\ref{sec:tps}), and a spatially varying aberration model (Sec.~\ref{sec:svamodel}).

\subsubsection{Source model}
\label{sec:learnable_source}

The source model combines an incident plane wave with the non-idealities in the system (e.g. a slow varying intensity modulation in the source or dust/scratches in the optical path). 
We learn a complex-valued source pattern that corrects for these non-uniformities across the input plane wave.

\subsubsection{Propagation model}
\label{sec:propagation_model}

We use angular spectrum method (ASM) propagation~\cite{goodman2005fourieroptics} to model free-space light transport. To include global aberrations into the propagation model, we incorporate a learned complex pupil function.

\subsubsection{Phase and amplitude SLM models}
\label{sec:phase_amplitude_slm_models}
Generating a hologram starts with the voltage values sent to the SLMs, represented by 8-bit digital input values.
These values for the phase and amplitude SLMs are passed through a learned lookup table (LUT), mapping digital input to phase and amplitude for their respective SLMs.

The phase LUT is parameterized by 256 coefficients corresponding to the 8-bit input values.
The amplitude LUT uses a modification of the phase LUT parameterization. 
Amplitude modulation is periodic over the full actuation range of 256 coefficients.
We select a contiguous region of 100 coefficients corresponding to one cycle of monotonic decreasing amplitude modulation from full transmission to full attenuation.
For variable resolution amplitude modulation, we bin pixels by holding digital input values constant over all of the pixels in a binning region.
Both LUTs are made differentiable by using 1D interpolation between coefficients.

Liquid-crystal-based SLMs have a characteristic known as field fringing (cross-talk), where actuation at a pixel affects neighboring pixels due to the transitions at pixel boundaries.
This cross-talk between pixels is modeled with a learnable convolution kernel.
Our model upsamples modulation values $2\times$ in horizontal ($x$) and vertical ($y$) directions to account for sub-pixel effects, with the cross-talk kernel being 5 pixels in upsampled space.

The phase with field fringing and the amplitude with field fringing are then converted to electric fields.
For the amplitude modulation, we assume uniform phase, and for the phase modulation, we assume uniform amplitude.
This approach allows us to model the amplitude modulation separately, providing a more accurate representation of the system.

\subsection{Calibration: mapping the complex field between modulators}
\label{sec:tps}
\begin{figure}[!t]
  \includegraphics[width=\textwidth]{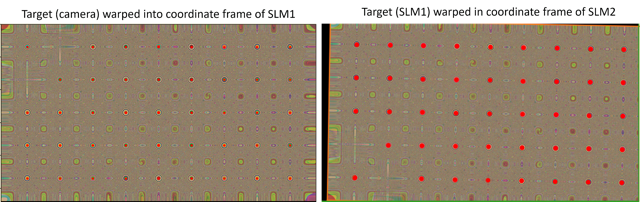}
  \caption{
  \textbf{Spatial light modulator (SLM) and camera alignment.}
  The left image shows a visualization of a grid of dots (asymmetric) reprojected into the coordinate system of the first spatial light modulator (SLM).
  The captured dots align with the pattern that was displayed on the first SLM (SLM1) to create these dots.
  The second image shows the Fresnel pattern after the transformation from the first SLM (SLM1) into the frame of the second SLM (SLM2).
  We also reproject the dots from the image capture back into the frame of the second SLM, which overlays well.
  }
\label{fig:tps_visualization}
\end{figure}

Generating high-quality images with a dual spatial light modulator (SLM) system requires knowledge of pixel-level alignment between modulators. We achieve this by establishing a deformation map between the two SLMs, referred to as SLM1 and SLM2.
This involves transforming the complex field from SLM1 into the coordinate system of SLM2.
This transformation is done with the thin plate spline (TPS) model using the implementation provided by Kornia~\cite{riba2019kornia}.

We first optimize a pattern for each SLM to display a grid of dots, which are displayed and captured sequentially on the camera.
We use an asymmetric pattern of dots to account for orientation changes caused by the beamsplitter and the 4$f$ systems.
After identifying the centers of these dots within the camera's coordinate system, we fit a TPS model from SLM1 to the camera and another from SLM2 to the camera.
We then apply linear algebra principles to calculate the TPS transformation between SLM1 and SLM2.
We further refine the parameters of this transformation through gradient descent, along with other parameters within our computational model.

The transformation process between SLM1 and SLM2 is illustrated in Fig.~\ref{fig:tps_visualization}.
Due to minor alignment issues, we cannot use the full field of view of the SLMs.
Therefore, our image quality evaluations are limited to areas where both SLM1 and SLM2 effectively modulate the wavefront.

\subsection{Calibration: modeling spatially varying aberrations}
\label{sec:svamodel}
Spatially varying aberrations present a significant challenge in holography, as they can compromise the quality of the displayed images in 2D and 3D.
These aberrations can result in significant disparities between the reconstructed images and their intended targets.
Traditional models often assume a linear shift-invariant (LSI) system, with a consistent point spread function across the field of view (FOV). 
However, this assumption does not hold true for many optical systems, which have spatially varying point spread functions, necessitating a more complex model. In our system we have two 4$f$ relays and other system non-idealities (e.g. shifts in aperture placement in the Fourier plane) which cause spatially varying aberrations.

We tackle spatially varying aberrations in holographic displays by utilizing a patch-wise method.
Each patch in the display incorporates a learned aperture function, allowing for localized aberration correction.
This method enhances image quality by addressing aberrations at the local level in the FOV.
We first patch the complex field into a grid of 18 x 24 of aforementioned aperture functions.
Each spatial aberration function is represented by a 256$\times$256 complex-valued function, which is resized to the resolution of the patch using bilinear interpolation in the complex domain using phase and amplitude respectively.
We then perform patch-wise convolution for each patch using the corresponding aperture function in Fourier space.
Afterwards we blend the patches together to model the effect of the spatially varying aberrations over the full FOV. 
In some experimental captures, the boundaries between aberration patches are visible. 
CITL is able to correct for most of the boundary effects that were not successfully blended.

As there are two 4$f$ relays in our dual-SLM system, we learn separate spatially varying pupil functions for each relay. We also include a global learned aperture for each relay.

\subsection{Calibration: pipeline}

\subsubsection{Calibration datasets}

We create calibration images by first optimizing random phase holograms (with random phase initialization) from a dataset of natural images.
We generate the phase SLM patterns by optimizing with an ideal ASM propagation model (Sec.~\ref{sec:propagation_model}).
The primary advantage of using pre-optimized patterns is that we can see image features during the software calibration.
An alternative would be to use speckle patterns where random phase is displayed on the hologram as proposed in Multisource Holography~\cite{gracemultisource23}.
This method is expected to work equally well, but the absence of any interpretable image features makes it challenging to debug the calibration process for phase SLMs.
For the amplitude SLM we use a combination of low-frequency patterns (e.g. constant, large blocks, and stripes) and high-frequency patterns (e.g. speckle patterns).

\subsubsection{Calibration procedure}
\label{sec:calibration_procedure}
To learn the parameters of our dual SLM model, we use a three-step calibration procedure with gradient descent in PyTorch.
For each calibration step, we capture an experimental dataset of images corresponding to input SLM patterns.

We first learn the parameters corresponding to the phase SLM using a dataset of 330 pre-computed random phase patterns. The amplitude SLM is held constant in this dataset, and during training, only parameters corresponding to the phase SLM are updated.

We then learn the parameters corresponding to the amplitude SLM using a dataset of 500 patterns, consisting of a combination of low-frequency and high-frequency patterns. Similarly, the phase SLM is held constant in this dataset and only parameters corresponding to the amplitude SLM are updated during training.

Finally, we fine-tune all the parameters for the dual system, using a dataset of 50 pre-computed random phase and smooth amplitude patterns. The dual system patterns are optimized with the calibrated model with individually learned amplitude and phase components.

The three-step training process reaches convergence in approximately 8 hours on an Nvidia A6000 GPU.
All implementation of our calibration was done using HoloTorch~\cite{chakravarthula2022differentiable}.

\subsubsection{Hologram generation}
\label{sec:hologramgeneration}
After training our dual SLM model, we solve for the SLM patterns for amplitude and phase for each desired hologram.
The patterns are initialized with constant (smooth) amplitude for the amplitude SLM and random phase for the phase SLM.
Patterns are optimized with the constraint of being in the monotonic region for the amplitude SLM and phase wrapped for the phase SLM.
A dual pattern optimization for a 2D hologram is on the order of minutes on an Nvidia A6000 GPU.

To vary the resolution of the amplitude SLM, we initialize patterns at desired resolutions, then use a differentiable upsampling operation to make constant blocks formed from pixels at the same resolution as the phase SLM before propagating through the model.
In back-propagation, the learnable parameters in the amplitude SLM are the initial low resolution values.

\section{Active camera-in-the-loop (CITL)}
\label{sec:hologramcitl}
Active camera-in-the-loop (CITL)~\cite{peng2020neural} is an online calibration process that we use to fine-tune images, correcting for small model mismatch and artifacts. CITL adjusts the SLM patterns for one hologram based on live feedback from the camera. 

\begin{figure}[!t]
  \centering
  \includegraphics[width=0.85\textwidth]{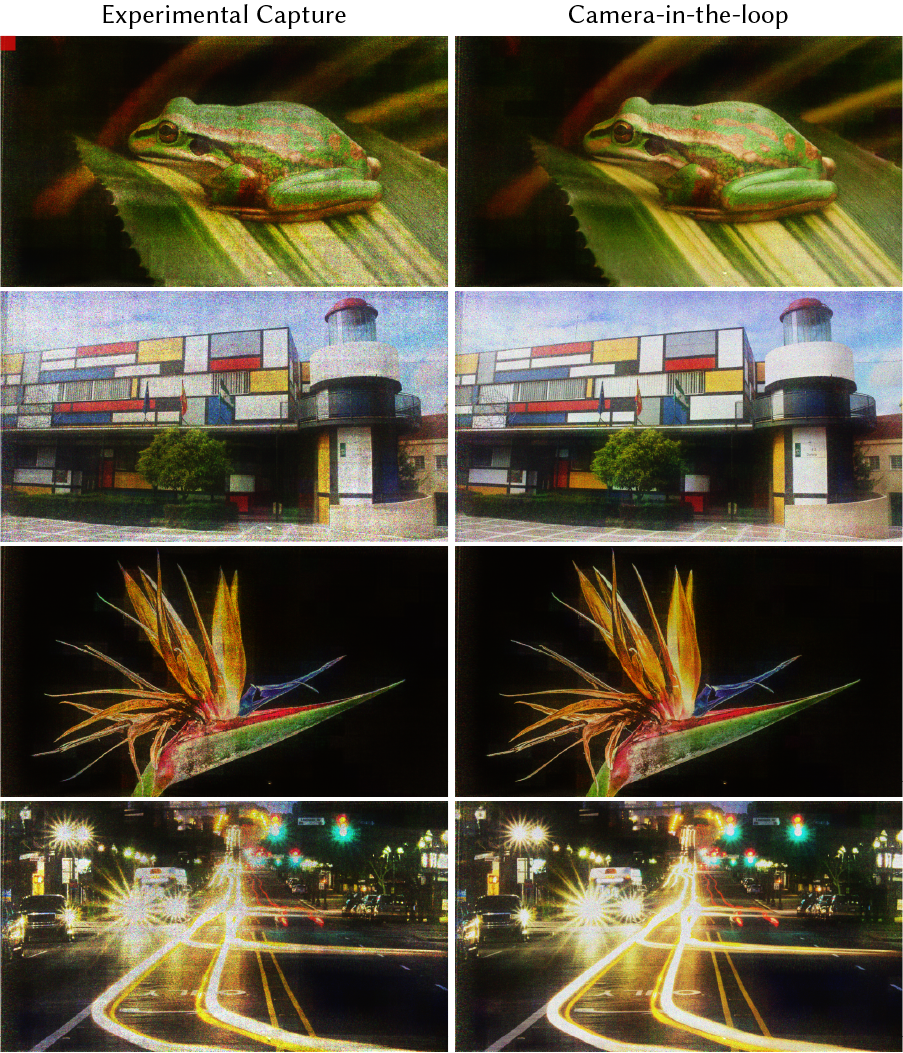}
  \caption{\textbf{Active camera-in-the-loop (CITL) comparison.} In the first column we show experimentally captured holograms optimized with our learned model but no active CITL. In the second column we show experimentally captured holograms that were iteratively fine-tuned based on camera feedback using CITL. Amplitude pixel size is $\Delta_a = 480$~\textmu m, shown in red in the top left image. Flower source image by Paul Longinidis (CC BY 2.0).}
  \label{fig:citl_comparison}
\end{figure}

We first optimize the amplitude and phase SLM patterns offline based on our learned calibration model (Sec.~\ref{sec:calibration}). 
Displaying these two patterns and capturing the resulting hologram gives our starting point for CITL. 
We use the transformation process of our TPS model (Sec.~\ref{sec:tps})
to transform the camera capture to the SLM space to do our optimization. This transformation gives us precise pixel-to-pixel alignment between the captured image and target, which allows for replacing the simulated model output used in our training loop with the transformed camera capture. We update the amplitude and phase SLM patterns for the hologram based on back-propagation with the transformed camera capture at each gradient descent step. Fig.~\ref{fig:citl_comparison} demonstrates the results of fine-tuning. CITL image quality improvement is clearly visible, as it reduces speckle, image artifacts, and improves color fidelity.

\section{Lowest resolution modulation}
\begin{figure}[!t]
  \includegraphics[width=\textwidth]{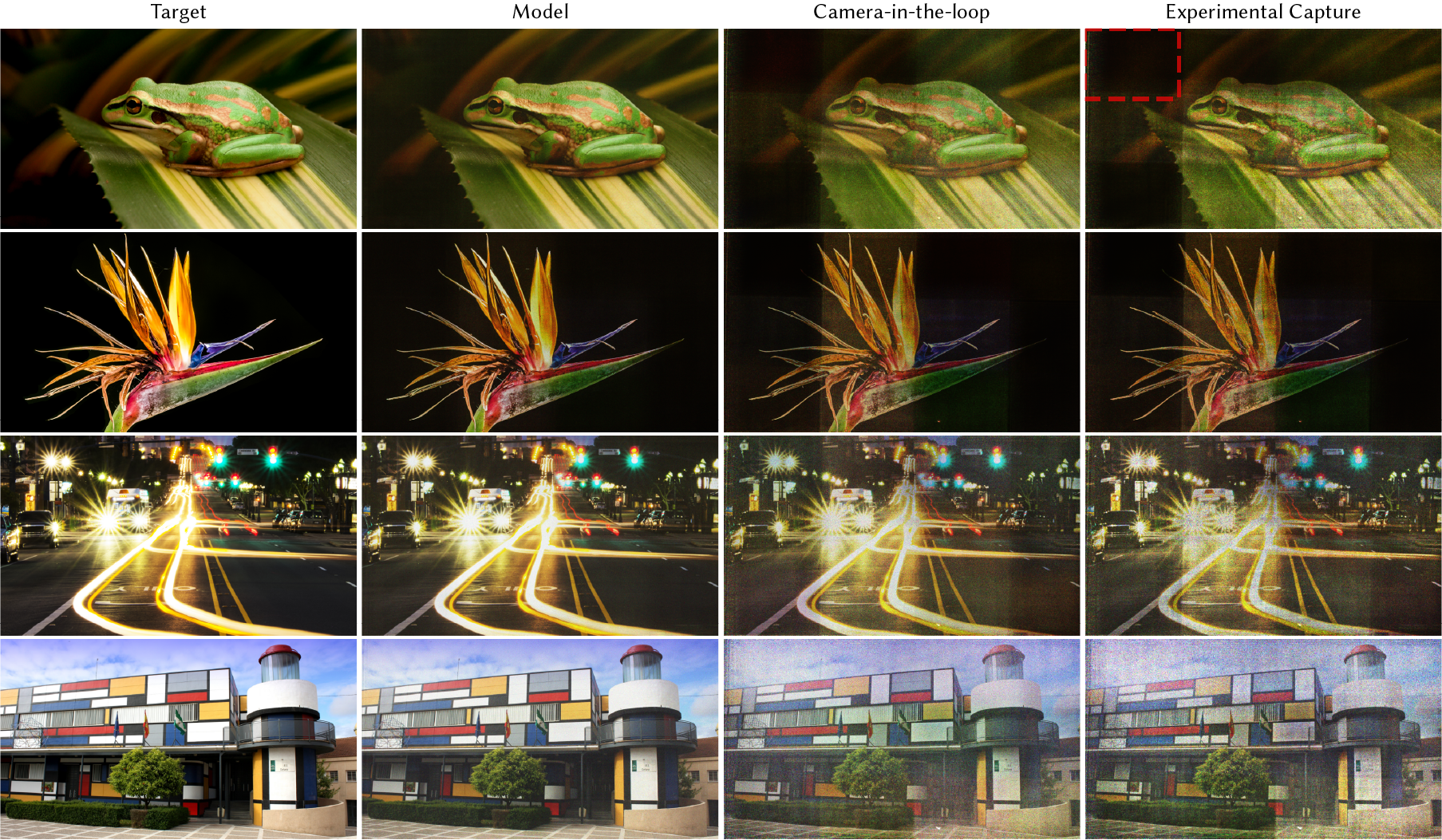}
  \caption{\textbf{Lowest resolution dual modulation.}
  At very low resolution amplitude modulation (amplitude pixel size $\Delta_a = 2880$~\textmu m), light is not fully redirected within the amplitude pixel region by the phase modulator, resulting in visible pixel boundaries in the experimental captures of holograms optimized with our learned model and the experimental captures with additional camera-in-the-loop (CITL) fine-tuning. The calibrated model output shows that the pixel boundaries are imperceptible in simulation, indicating potential for improvement in experimental suppression of pixel boundaries. 
 Amplitude pixel size is shown in red in the top right image. Flower source image by Paul Longinidis (CC BY 2.0).}
\label{fig:model_output_comparison}
\end{figure}

At the lowest resolution amplitude modulation, where amplitude pixel size is $\Delta_a = 2880$~\textmu m, individual pixel boundaries are visible in experimental captures. This resolution is beyond the recommended bound in the main text, so light cannot be fully redirected within the amplitude pixel region by the phase SLM. In Fig.~\ref{fig:model_output_comparison}, the pixel boundaries are visible in both experimental captures and experimental captures with CITL fine-tuning. CITL corrects slightly for some of the pixel boundary artifacts but cannot suppress all of the amplitude pixel structure. For images with high-frequency content, such as the Mondrian building, the pixel structure is less visible in experimental captures. The simulated outputs produced by our calibrated model do not contain any perceptible pixel boundaries. This suggests that with improvements to the calibration approach, it may be possible to further correct for the mismatch between experimental captures and simulated model outputs. Future work can move towards using extremely low resolution amplitude modulators with reduced pixel boundary artifacts.

\section{Amplitude patterns}
Here we include examples of the amplitude SLM patterns for both 2D images and focal stacks. 

\begin{figure}[!t]
  \includegraphics[width=\textwidth]{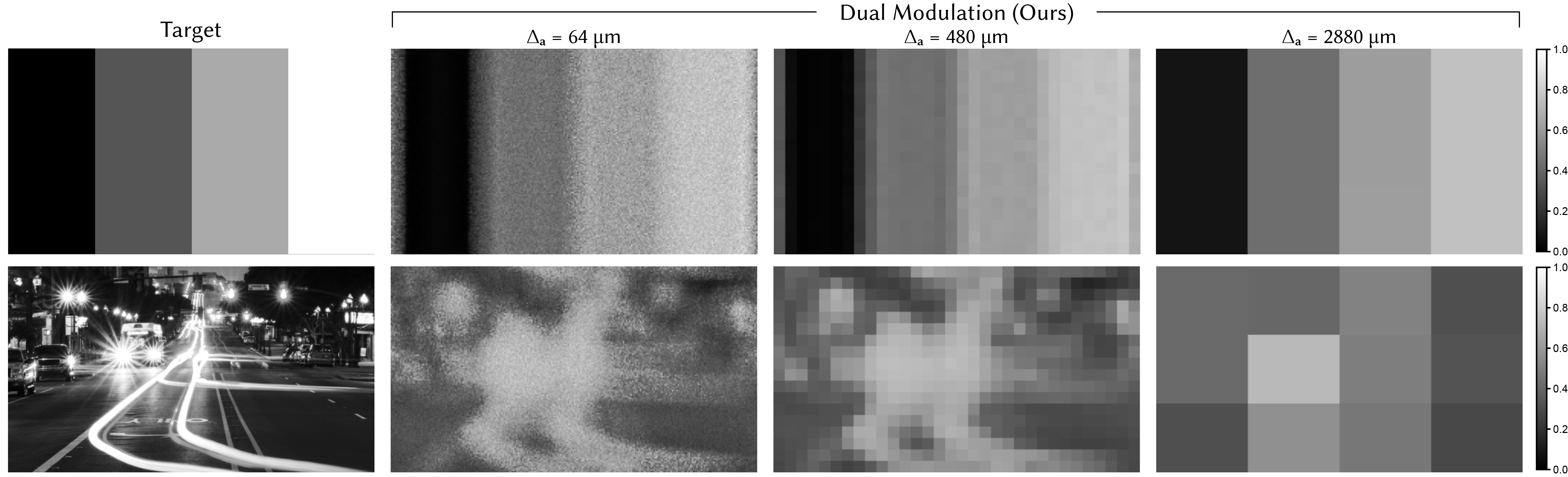}
  \caption{\textbf{2D image amplitude patterns.}
  Amplitude patterns for amplitude pixel sizes $\Delta_a = 64$~\textmu m, $\Delta_a = 480$~\textmu m, and $\Delta_a = 2880$~\textmu m. Patterns correspond to the simulation grayscale images in the main text. Full light transmission corresponds to 1.0 and full light attenuation corresponds to 0. The amplitude pattern resembles the low-frequency structure of the target.}
\label{fig:2Dimagepatterns}
\end{figure}

Fig.~\ref{fig:2Dimagepatterns} demonstrates amplitude patterns for the grayscale simulated images in the main text, for amplitude pixel sizes $\Delta_a = 64$~\textmu m, $\Delta_a = 480$~\textmu m, and $\Delta_a = 2880$~\textmu m. Full light transmission corresponds to 1.0 and full light attenuation corresponds to 0. The amplitude pattern resembles the low-frequency structure of the target.

\begin{figure}[!t]
  \includegraphics[width=\textwidth]{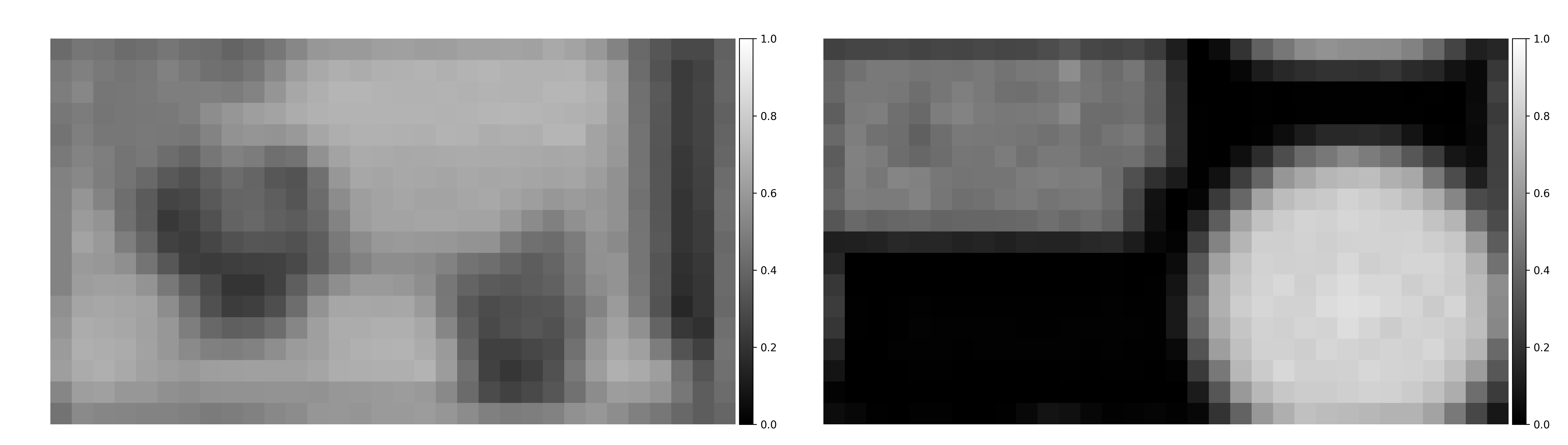}
  \caption{\textbf{Focal stack amplitude patterns.}
  Amplitude patterns for amplitude pixel size $\Delta_a = 480$~\textmu m. Patterns are for the natural scene color focal stack (left) and the grayscale focal stack (right) from the main text. Full light transmission corresponds to 1.0 and full light attenuation corresponds to 0. In general, the amplitude pattern resembles the low-frequency structure of the target.}
\label{fig:focalstackpatterns}
\end{figure}

Fig.~\ref{fig:focalstackpatterns} demonstrates amplitude patterns for the focal stacks in the main text, with amplitude pixel size $\Delta_a = 480$~\textmu m. The left pattern is for the green channel of the natural scene color focal stack and the right pattern is for the grayscale focal stack containing text and a Siemens star. Full light transmission corresponds to 1.0 and full light attenuation corresponds to 0. Intuitively, the amplitude pattern resembles the low-frequency structure of the target, attenuating light in dark regions and transmitting light in bright regions.
As the intensities of objects do not vary significantly through defocus, one amplitude pattern captures the structure across the focal stack.

\newpage 
\clearpage 

\section*{Acknowledgments}
Flower source image by Paul Longinidis (CC BY 2.0) is available at~\url{https://www.flickr.com/photos/128235612@N06/18033751041/}. Image was cropped but otherwise unmodified.
Houses source image by Madeleine Deaton (CC BY 2.0).